# Suspended waveguide-enhanced near-infrared photothermal spectroscopy for ppb-level molecular gas sensing on a chalcogenide chip

Kaiyuan Zheng[1,2,†], Hanyu Liao[1,2,†], Fengbo Han[3,†], Xueying Wang[4], Yan Zhang[3], Jiaxin Gu[3], Pengcheng Zhao[1,2], Haihong Bao[1,2], Shaoliang Yu[3], Qingyang Du[3*], Lei Liang[5], Chuantao Zheng[4*], Wei Jin[1,2*], and Lijun Wang[5]

[1]Department of Electrical and Electronic Engineering and Photonics Research Institute, The Hong Kong Polytechnic University, Hong Kong, 999077, China

[2]Photonics Research Center, The Hong Kong Polytechnic University Shenzhen Research Institute, Shenzhen, 518071, China

[3]Zhejiang Lab, Hangzhou, 311121, China

[4]State Key Laboratory of Integrated Optoelectronics, JLU Region, College of Electronic Science and Engineering, Jilin University, Changchun, 130012, China

[5]State Key Laboratory of Luminescence and Applications, Changchun Institute of Optics, Fine Mechanics and Physics, Chinese Academy of Sciences, Changchun 130033, China

[†]These authors contributed equally to this work

[*]Corresponding authors: qydu@zhejianglab.edu.cn, zhengchuantao@jlu.edu.cn, wei.jin@polyu.edu.hk

**Abstract:**

On-chip waveguide sensors have attracted significant attention recently due to their potential for high level integration. However, so far on-chip gas sensing based on traditional laser absorption spectroscopy has demonstrated low detection sensitivity, due to weak light-gas interaction over a limited interaction distance. On-chip photothermal spectroscopy (PTS) appears to be a powerful technique to achieve higher sensitivity, its performance is yet constrained to parts-per-million (ppm)-level due to small fraction of evanescent field in the light-gas interaction zone and fast thermal dissipation through the solid substrate. Herein, we demonstrated suspended chalcogenide glass waveguide (ChGW)-enhanced PTS that overcomes these limitations, enabling highly sensitive parts-per-billion (ppb)-level molecular gas sensing. We fabricated a nanoscale suspended ChGW with low loss of 2.6 dB/cm using CMOS-compatible two-step patterning process. By establishing an equivalent PTS model to guide the optimization of the ChGW geometry, we achieved a 4-fold increase in the absorption-induced heat source power and a 10.6-fold decrease in the equivalent heat conductivity, resulting in a 45-fold enhancement in photothermal phase modulation efficiency over the non-suspended waveguides. Combining with a high-contrast waveguide facet-formed Fabry-Perot interferometer, we achieved an unprecedented acetylene detection limit of 330 ppb, a large dynamic range close to 6 orders of magnitude, and a fast response of less than 1 s. The overall system exhibits a noise-equivalent absorption coefficient of $3.8\times10^{-7}$ $cm^{-1}$, setting a new benchmark for photonic waveguide gas sensors to the best of our knowledge. This work provides a key advancement towards prototyping an integrated sensor-on-a-chip for highly sensitive and background-free photonic sensing applications.

## 1. Introduction

With the increasing demand of portable sensors for environmental monitoring and wearable healthcare,[1,2] on-chip photonic gas sensors have garnered significant attention due to their advantages in size, weight, power, and cost (SWaP-C).[3-5] Most on-chip gas sensors reported so far are based on direct absorption spectroscopy (DAS), relying on the interaction of gas molecules with the evanescent field of a photonic waveguide[6-9]. They operate in the near-infrared (NIR) or mid-infrared (MIR) regions and have waveguide length on the order of centimeters, achieving detection sensitivity from hundreds to several ppm[10,11]. An alternative approach leverages a highly sensitive on-chip resonant cavity and detects the refractive index (RI) change in the presence of gas molecules. This method achieves sensitivity down to ppm level but lacks specificity.[12,13] The limited sensitivity is due to the intrinsically short optical pathlength, residual fringing noise, and weak light-gas interaction particularly in the NIR region.

Recent advents in photothermal spectroscopy (PTS) have proved itself to be a highly sensitive and selective technique for trace gas detection.[14,15] Gas molecules absorb a pump beam with wavelength tuned to a specific gas absorption line, which generates heat through molecule collisions and changes the temperature distribution of the surrounding medium through heat conduction. PTS measures photothermally induced RI change by detecting the phase modulation of probe beam, benefitting from both the high sensitivity of RI sensing and the high specificity of absorption spectroscopy.[16-18] PTS is recognized as a background-free technique, where phase measurement offers superior sensitivity, larger dynamic range, and reduced susceptibility to fringing noise compared to intensity-based DAS.[15] In recent years, PTS has transitioned from free space systems with millimeter-scale mode field diameters (MFD) to fiber-optic platforms with micrometer-scale MFD and achieves significantly enhanced sensitivity due to improved photothermal (PT) efficiency, which is inversely proportional to the MFD. Nanometer-scale waveguides could have a much smaller MFD, which would enable a higher PT efficiency and have progressed from theoretical designs to experimental demonstrations recently.[19] However, their performance is limited by the small fraction of evanescent field for light-gas interaction, resulting in low level of heat source power due to absorption. In addition, rapid heat dissipation from the solid bottom-cladding material, such as silica ($SiO_2$) with high thermal conductivity, reduces heat accumulation and impairs PTS sensing performance. Prior studies show that using lower-thermal-conductivity polymer bottom-cladding[20] or introducing air gap[21] can mitigate heat leakage and improve light-gas overlap. However, there remains a lack of systematic guidance and quantitative evaluation on how to optimize these approaches for gas sensing, particularly regarding phase modulation amplification and thermal resistance. The relative performance of the suspended method compared to other thermal isolation strategies also remains unclear.

Herein, we present an on-chip suspended waveguide-enhanced PTS (SWE-PTS) to address the limitations of current photonic sensors. We theoretically establish an equivalent PT model to quantitatively evaluate the heat generation, conduction dynamics and PT modulation on a suspended ChGW surrounded by absorptive gas, which guides the optimization of the waveguide material, geometry and length to achieve optimal PT efficiency. Theoretical analysis reveals that the suspended ChGW demonstrates a nearly 2 orders of magnitude enhancement in PT efficiency than traditional non-suspended one. We then fabricated a low-loss suspended ChGW with 3-μm-thick air buffer zone (bottom cladding) by using a standard CMOS-compatible two-pattern membrane release process. Compared with solid $SiO_2$ bottom cladding, the air buffer zone significantly enhanced the evanescent field, resulting in a larger optical absorption and thus higher heat source power. The thermal conductivity of air is more than 50 times smaller than that of $SiO_2$, which promotes heat accumulation in the buffer zone and hence induces a large temperature variation. Such enhanced heat accumulation, coupled with the large thermos-optic coefficient of the ChGW material, enables significant enhancement of PT phase modulation and hence higher gas detection sensitivity. We further validate the capability of its gas molecular sensing capability of SWE-PTS by incorporating an on-chip Fabry-Perot interferometer utilizing the inherent reflections at the waveguide facets, enabling an integrated device for heat generation and detection simultaneously. As a result, our photonic waveguide gas sensor achieves unprecedented ppb-level gas detection for the first time in the NIR, which is particularly significant given the orders of magnitude weaker absorption in the NIR as compared to MIR, and a short chip length of only 1.2 cm. Photonics in the NIR is relatively mature and easy to integrate with light sources and detectors, paving the way for next-generation high-sensitivity, high-specificity, and large-scale integrated photonic sensors.

## 2. Principle

**2.1 Theoretical formulation.** The suspended waveguide uses air as the top and bottom cladding, a ridge structure as the core layer and a doped silicon (Si) as the substrate, as shown in Fig. 1(a). A pump-probe configuration is used, where a wavelength-modulated pump beam, with its nominal wavelength tuned to a gas absorption line, propagates along the suspended ChGW with evanescent field extending into the air cladding. Gas molecules absorb the pump energy in the evanescent field, resulting in heat generation due to the non-radiative relaxation. The corresponding heat source $Q(x, y, t)$ follows the intensity profile of the pump evanescent field in the air region. Since the pump wavelength is modulated around the gas absorption line, the heat generation is periodic and can be decomposed into a series of harmonics of the modulation frequency. The $n^{\text{th}}$ harmonic heat source may be expressed as:

$$Q_{\mathrm{n}}(x, y, t) = H_{\mathrm{n}} \alpha C I_{\mathrm{p}}^{\mathrm{gas}}(x, y) e^{\mathrm{j} n \omega t} \quad (1)$$

where $H_{\mathrm{n}}$ is a harmonic coefficient depending on the depth of wavelength-modulation with respect to the gas absorption lineshape function,[14] $\alpha$ the peak absorption coefficient of the analyte (for $C_2H_2$, $\alpha$=1.05 cm$^{-1}$ at 1531.58 nm), $C$ the gas concentration, $\omega = 2\pi f$ with $f$ representing the pump modulation frequency. $I_{\mathrm{p}}^{\mathrm{gas}}(x, y)$ is the pump intensity distribution with superscript denotes the light intensity in the gas region. Taking the $2^{\text{nd}}$ harmonic heat source as an example, $n$=2, $H_2$=0.343, which is obtained with optimal modulation depth of 2.2 times of the absorption linewidth.[22] The total heat source power $P_{\mathrm{Q}}$ due to absorption of the incident pump power $P_{\mathrm{p}}$ may be written as:

$$P_{\mathrm{Q}}(t) = \iint Q_2 \, dxdy = H_2 \alpha C P_{\mathrm{p}} \Gamma e^{\mathrm{j} 2 \omega t} \quad (2)$$

$$\Gamma := \frac{\iint I_{\mathrm{p}}^{\mathrm{gas}} dxdy}{P_{\mathrm{p}}} = \frac{n_{\mathrm{gp}} \iint_{\mathrm{gas}} \varepsilon \left| E_{\mathrm{p}} \right|^2 dxdy}{n_{\mathrm{gas}} \iint_{\mathrm{total}} \varepsilon \left| E_{\mathrm{p}} \right|^2 dxdy} \quad (3)$$

here $\Gamma$ is the light-gas confinement factor that scales the interaction strength,[23] $n_{\mathrm{gp}}$ is the group index of the pump mode in the suspended ChGW, $n_{\mathrm{gas}}$ is the RI of gas, $\varepsilon(x,y)$ and $E_{\mathrm{p}}(x,y)$ are respectively the permittivity and the electric field distribution of the pump. According to Eq. (2-3), the incident pump power $P_{\mathrm{p}}$ is converted to the heat source power $|P_{\mathrm{Q}}|$ scaled by the unit-less factor $\Gamma$, which benefits from the pump group index $n_{\mathrm{gp}}$ and the fraction of the pump evanescent field $\gamma_{\mathrm{p}}$. The suspended design of ChGW significantly improves the light-gas interaction through a higher evanescent field in gas, hence a higher heat power $P_{\mathrm{Q}}$.[24,25]

The heat source $Q$ leads to a change of temperature distribution $\Delta T(x,y,t)$ across the waveguide cross-section through heat conduction, which can be detected by a probe beam in the form of phase modulation. Fig. 1(b) shows the resulting temperature change for the pump modulation frequency of $f$ = 1 kHz and $n$ = 2. The suspended design creates a thermal buffer zone before the heat reaches the Si substrate, which enables heat accumulation around the probe mode and hence a larger PT phase modulation. To quantify the improvement of heat accumulation by suspended design, we introduce an equivalent heat transfer model by decomposing the temperature change $\Delta T$ into a product:

$$\Delta T(x,y,t) = T_{\mathrm{eq}}\psi_{\mathrm{T}}(x,y)e^{\mathrm{j}2\omega\mathrm{t}} \tag{4}$$

where $\psi_{\mathrm{T}}$ is the temperature spatial mode profile capturing the geometric dependence of the temperature change, $T_{\mathrm{eq}}$ is the equivalent temperature amplitude scaling the mode profile. This separable form arises since the fixed de-modulation frequency and fixed pump wavelength define a specific mode $\psi_{\mathrm{T}}(x,y)$ for the heat transfer in waveguide cross-section. For SWE-PTS, we are only interested in the overlap between the probe field $\psi_{\mathrm{b}}$ and the temperature field $\psi_{\mathrm{T}}$. Thus, we normalize $\psi_{\mathrm{T}}$ by projecting it onto the probe mode under a weighted inner product:

$$\iint_{\mathrm{total}} \psi_{\mathrm{b}}^{*}\psi_{\mathrm{T}}\psi_{\mathrm{b}}\, dxdy \coloneqq 1 \tag{5}$$

here $\psi_{\mathrm{b}}$ is the probe mode field. Eq. (5) ensures unit coupling strength between $\psi_{\mathrm{T}}$ and the PT phase modulation on mode $\psi_{\mathrm{b}}$. Based on Eq. (4-5), the heat transfer can be analytically solved in integration form for constant temperature boundaries (Supplementary Note 1). This approach permits analysis of the system's macroscopic behaviour without resolving spatial transients, i.e. the equivalent temperature amplitude $T_{\mathrm{eq}}$:

$$T_{\mathrm{eq}} = \iint_{\mathrm{total}} \psi_{\mathrm{b}}^{*}\tilde{T}\psi_{\mathrm{b}}\, dxdy = \frac{P_Q}{j2\omega_m C_{eq} - \kappa_{eq}} \tag{6}$$

where we have introduced equivalent volumetric heat capacity $C_{eq}$ and heat conductivity $\kappa_{\mathrm{eq}}$ to conclude the whole heat transfer characteristics:

$$\kappa_{\mathrm{eq}} = \iint_{\mathrm{total}} \nabla\cdot(\kappa(x,y)\nabla\psi_{\mathrm{T}})dxdy \tag{7}$$

$$C_{eq} = \iint_{total} C(x,y)\psi_T dxdy \tag{8}$$

According to Eq. (7-8), $C_{eq}$ and $\kappa_{\mathrm{eq}}$ captures the overall thermal convection and conduction behavior in the x-y plane experienced by $\psi_T$. The equivalent temperature $T_{\mathrm{eq}}$ is actually the temperature value probed by the probe mode $\psi_{\mathrm{b}}^{*}\psi_{\mathrm{b}}$. To achieve maximum phase modulation efficiency, we choose modulation frequency $\omega_m = 1kHz \ll C_{eq}/2\kappa_{eq}$ to obtain maximum $T_{\mathrm{eq}} = -P_Q/\kappa_{eq}$, and the error of this approximation is only 0.5% in our case. Hence, the $T_{eq}$ depends on the core-layer material thermal conductivity and the suspended ChGW geometric parameters. For molecular gas sensing, we measure the phase modulation $\Delta\varphi_{\mathrm{b}}$ on the probe beam induced by the temperature change through thermal-optic effect (TOE), which could be expressed as:

$$\Delta\varphi_{\mathrm{b}} \approx -\frac{2\pi n_{\mathrm{gb}}L}{\lambda_{\mathrm{b}}}\frac{P_{\mathrm{Q}}}{\kappa_{\mathrm{eq}}}\left(e_{\mathrm{ChG}}^{\mathrm{TO}}\iint_{\mathrm{ChG}}\psi_{\mathrm{b}}^{*}\psi_{\mathrm{T}}\psi_{\mathrm{b}}\,dxdy + e_{\mathrm{gas}}^{\mathrm{TO}}\iint_{\mathrm{gas}}\psi_{\mathrm{b}}^{*}\psi_{\mathrm{T}}\psi_{\mathrm{b}}\,dxdy\right)e^{\mathrm{j}2\omega\mathrm{t}} \tag{9}$$

here $\lambda_{\mathrm{b}}$ is the probe wavelength, $n_{\mathrm{gb}}$ the group index of the probe mode, $L$ the waveguide length, and $e^{\mathrm{TO}} = dn/dT$ is the TO coefficient with subscript ChG and gas denote the core-layer ChG and outside gas. According to Eq. (9), the PT phase modulation is proportional to $\kappa_{\mathrm{eq}}^{-1}$. As compared to the conventional waveguide employing $SiO_2$ as the bottom cladding, the suspension design reduces the equivalent thermal conduction $\kappa_{\mathrm{eq}}$ and hence a much higher heat accumulation quantified by $T_{\mathrm{eq}}$.

To compare the SWE-PTS among different suspended geometric parameters, including the rib width ($w$), height ($h$) and rib ratio ($k$), we define a normalized PT phase modulation efficiency:

$$k^{*}(w,h,k) \coloneqq \frac{\Delta\varphi_{\mathrm{b}}}{\alpha C P_{\mathrm{p}} L} \propto \Gamma\frac{e_{\mathrm{eq}}^{\mathrm{TO}}}{\kappa_{\mathrm{eq}}} \tag{10}$$

here $k^{*}(w,h,k)$ is defined as the normalized PT phase modulation efficiency, $e_{\mathrm{eq}}^{\mathrm{TO}}$ represents the terms in the parenthesis in Eq. (10) and is regarded as the equivalent thermo-optic coefficient. According to Eq. (10), the suspended waveguide achieves superior PT phase modulation efficiency through dual optimization mechanisms: optical enhancement via enlarged optical absorption ($\Gamma$), and thermal enhancement via combining larger equivalent thermal resistance ($1/\kappa_{\mathrm{eq}}$) and amplified thermo-optic response ($e_{\mathrm{eq}}^{\mathrm{TO}}$). Numerical calculation reveals that suspending a ChG core-layer with a bottom-cladding thickness of several microns can improves the $k^{*}$ value by nearly 2 orders of magnitude. This remarkable improvement primarily stems from a 3–5 times increment in $\Gamma$ and 10–30 times larger $e_{\mathrm{eq}}^{\mathrm{TO}}/\kappa_{\mathrm{eq}}$, making it possible for highly sensitive gas detection on a short-length chip. More details of SWE-PTS refer to Supplementary Note 1.

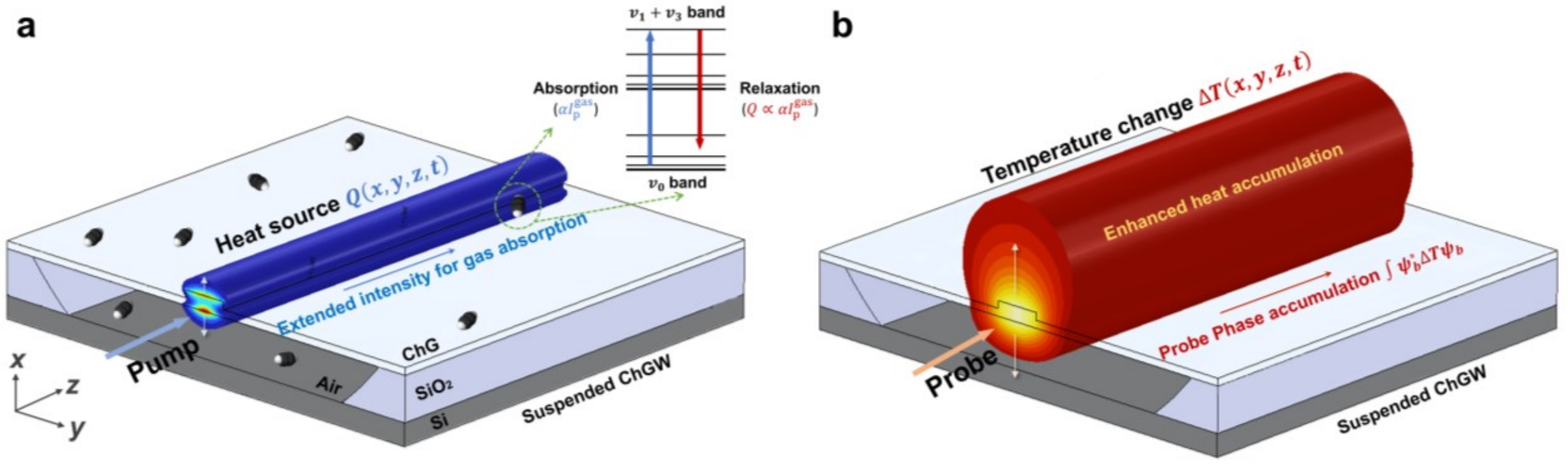

**Fig. 1 Principle of the SWE-PTS. (a)** Heat generation by pump absorption. The pump beam exhibits a strong evanescent field in the gas region above and below the waveguide. The gas molecules absorb the pump energy and transit to higher energy level, which subsequently returns to the ground state via thermal relaxation and produces heat source $Q$. The profile of the heat source follows the evanescent pump intensity profile $\psi_{\mathrm{p}}^{*}\psi_{\mathrm{p}}$ as a two-petal shape in the air. The suspended waveguide design enhances the product of fractional power in the evanescent field and the mode group index, which is referred as optical enhancement. The inset shows the exemplary energy levels of acetylene for the heat generation process. **(b)** Temperature induced probe phase modulation. Due to the heat conduction, the temperature distribution $\psi_{\mathrm{T}}$ is no longer of a two-petal shape, where the $T$ is approximately uniform in the ChG cross-section and it extends and diminishes in the air due to their large differential thermal conductivity. The heat accumulation is improved by the suspended design, which prevents the rapid heat loss through the ChG and substrates. The probe beam experiences phase change with respect to the spatial integration of optical and thermal mode field $\psi_{\mathrm{b}}^{*}\psi_{\mathrm{T}}\psi_{\mathrm{b}}$. The suspended design reduces the heat conduction and improves the thermo-optic effect as $e_{\mathrm{eq}}^{\mathrm{TO}}/\kappa_{\mathrm{eq}}$, which is referred as thermal enhancement.

**2.2 Optimization of suspended ChGW.** The waveguide material and geometry work together to determine the optical and thermal properties and hence the SWE-PTS performance. The selection criteria of the material, especially the core-layer, include large TOE, large thermal expansion effect (TEE) , and low $\kappa$ coefficient. To facilitate comparison, we use a PT figure-of-merit (FOM) of $n\times(\mathrm{TOE}+\mathrm{TEE})/\kappa$.[26,27] ChG exhibits an FOM of $2.87\times10^{-4}$ m/W, which is 2–3 orders of magnitude larger than Si, silicon nitride ($Si_3N_4$), and lithium niobate ($LiNbO_3$).[20] Therefore, ChG is selected as the core-layer material in our suspended waveguide design. Specifically, we choose the ChG composition $Ge_{28}Sb_{12}Se_{60}$, also commercially known as IG5, as it is resistant to both photo-darkening effect and oxidation in ambient environment compared with other arsenic-containing compositions.

To optimize the suspended ChGW structure, we compare the phase modulation efficiency $k^*$ for different waveguide parameters using COMSOL Multiphysics. As shown in Figs. 2(a)-(e), the optimal parameters are $w$=1000 nm, $h$=300 nm, $k$=0.9, and $h_2$=10 μm. Considering that a thin slab with a large rib ratio $k= h_1/h$ is mechanically fragile which may lead to collapse of the entire suspended structure, we decide to select the set of $w$=1000 nm, $h$=300 nm, $k$=0.5, and $h_2$=3 μm for a robust waveguide design without compromising too much of its performance. The waveguide length $L_{\mathrm{wg}}$ is determined by balancing the PT phase modulation with transmission loss. There is a trade-off, as the PT phase modulation $\Delta\varphi_{\mathrm{b}}$ generally increases with the waveguide length. However, a long waveguide diminishes the fringe contrast ($v$) of the waveguide interferometer due to increased transmission loss, and thus reduces the efficiency of phase-to-intensity conversion and the overall PTS signal. Fig. S3 shows the dependence of normalized $\Delta\varphi_{\mathrm{b}}$ and $v$ on the waveguide length. The variation of the PTS signal with the ChGW length is shown in Fig. 2(f), which indicates an optimal length of ~0.8 cm. In this setup, the actual ChGW length is 1.2 cm, which remains ~90% of the optimal value (Details in Supplementary Note 3).

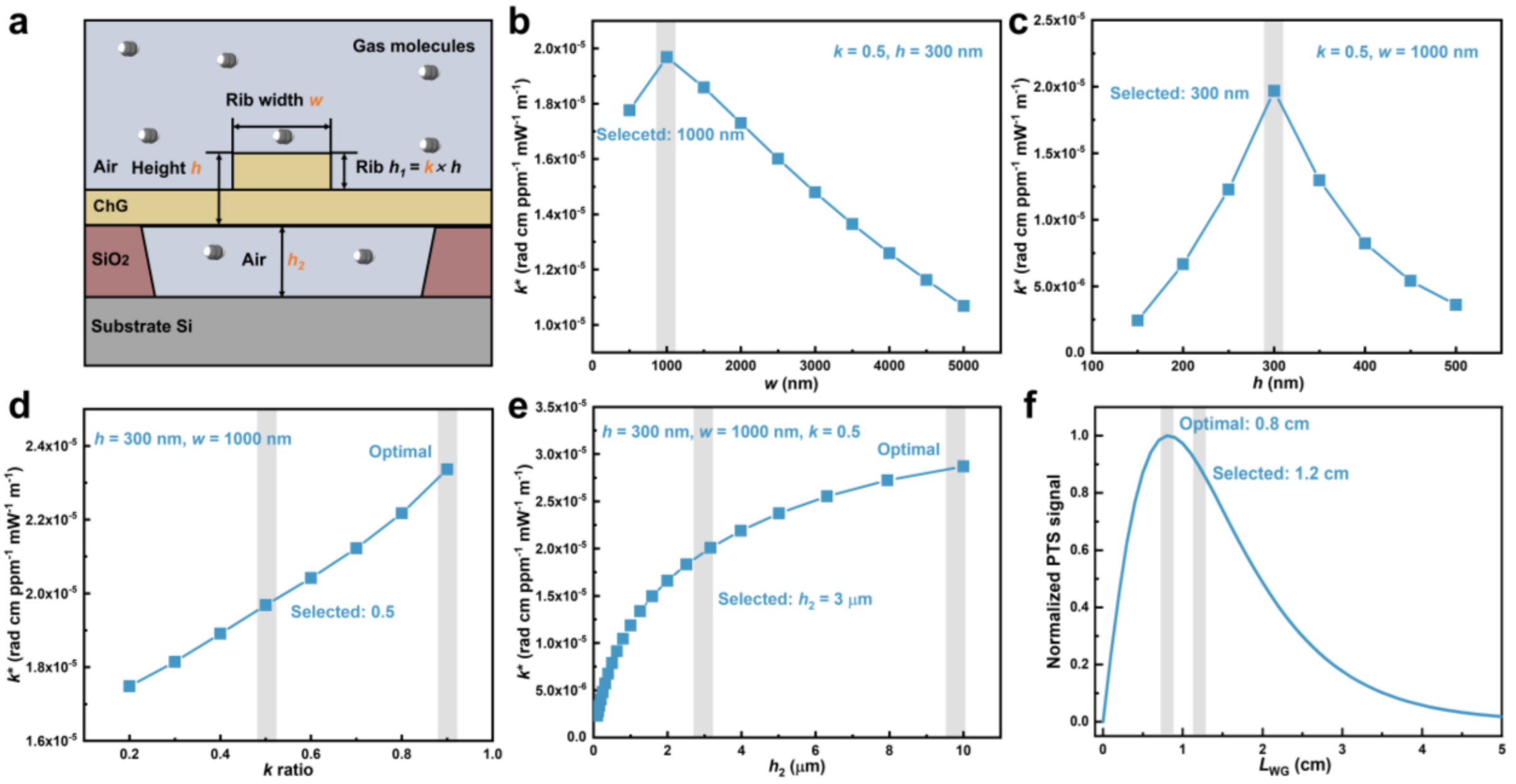

**Fig. 2 Optimization of the suspended ChGW for enhanced SWE-PTS. (a)** Schematic of the suspended waveguide with air serving as the top and bottom claddings. The phase modulation efficiency as function of **(b)** rib width (*w*), **(c)** core-layer height (*h*), **(d)** rib ratio (*k*), and **(e)** bottom buffer zone thickness ($h_2$). **(f)** The overall PTS signal for varying waveguide length.

**2.3 Comparison with non-suspended ChGW.** We then evaluated the optical and thermal enhancement in the optimized suspended ChGW over the non-suspended case through numerical simulation. Here the bottom-cladding is air for the suspended waveguide and $SiO_2$ for the non-suspended one.[28,29] Numerical calculation is performed with dimensions of $w$=1000 nm, $h$=300 nm, and $k$=0.5, with a 3-μm-thick air bottom-cladding as the buffer zone for suspended one, and $w$=1000 nm, $h$=300 nm for non-suspended strip one. Both TE and TM modes are supported in the suspended ChGW, but only the TM mode is selectively excited in this case due to its larger evanescent field, which would lead to stronger light-gas absorption. Fig. 3(a) shows the heat source due to gas absorption of the pump field ($\psi_p^*\psi_p$) and the probed thermal field ($T_{eq}\psi_b^*\psi_T\psi_b$) respectively, with their integrals correspond to the optical enhancement ($\Gamma$) and thermal enhancement ($e_{eq}^{TO}/\kappa_{eq}$) according to Eqs. (2) and (6), respectively. With a pump wavelength of ~1531 nm, the pump field in $TM_0$ mode exhibits a dual-sided evanescent field in air, resulting in a much higher $\Gamma$ of 95% than the non-suspended one of 24%.[30] For the thermal enhancement, the suspension design creates a thermal buffer zone, reduces the equivalent thermal conductivity $\kappa_{eq}$ from 0.793 to 0.075 W/(m·K) (Methods). Fig. 3(b) shows the dependence of the $\Gamma$, $e_{eq}^{TO}$, $\kappa_{eq}$ and $T_{eq}$ with the increasing thickness of air buffer zone $h_2$ from 0 to 5 μm. Compared to the non-suspended ChGW structure ($h_2$ = 0 μm), the values of $\Gamma$, $e_{eq}^{TO}$, $\kappa_{eq}$ and $T_{eq}$ for the case of 3-μm-thick buffer zone are enhanced by a factor of 4, 1.08, 10.6 and 42, respectively.

To illustrate the improvement in the PT response, we calculate the time-domain temperature modulation at the demodulation frequency of 2 kHz ($n$ = 2) and the results are shown in Fig. 3(c). The probed temperature change in the suspended structure increases from 0.33 mK to 13.7 mK, approximately 42 times that of the non-suspended one. This clearly shows that good thermal isolation promotes the rapid rise and fall of the thermal dynamic homeostasis of the PT effect.[14,31] The PT phase modulations of the suspended and non-suspended ChGW are determined by steady-state frequency-domain simulations are shown in Fig. 3(d). At a demodulation frequency of 2 kHz, a significant $k^*$ enhancement of 45 times is achieved, which evidences that the suspended structure a better option for PTS than the non-suspended one. Detailed comparisons are provided in Supplementary Note 4.

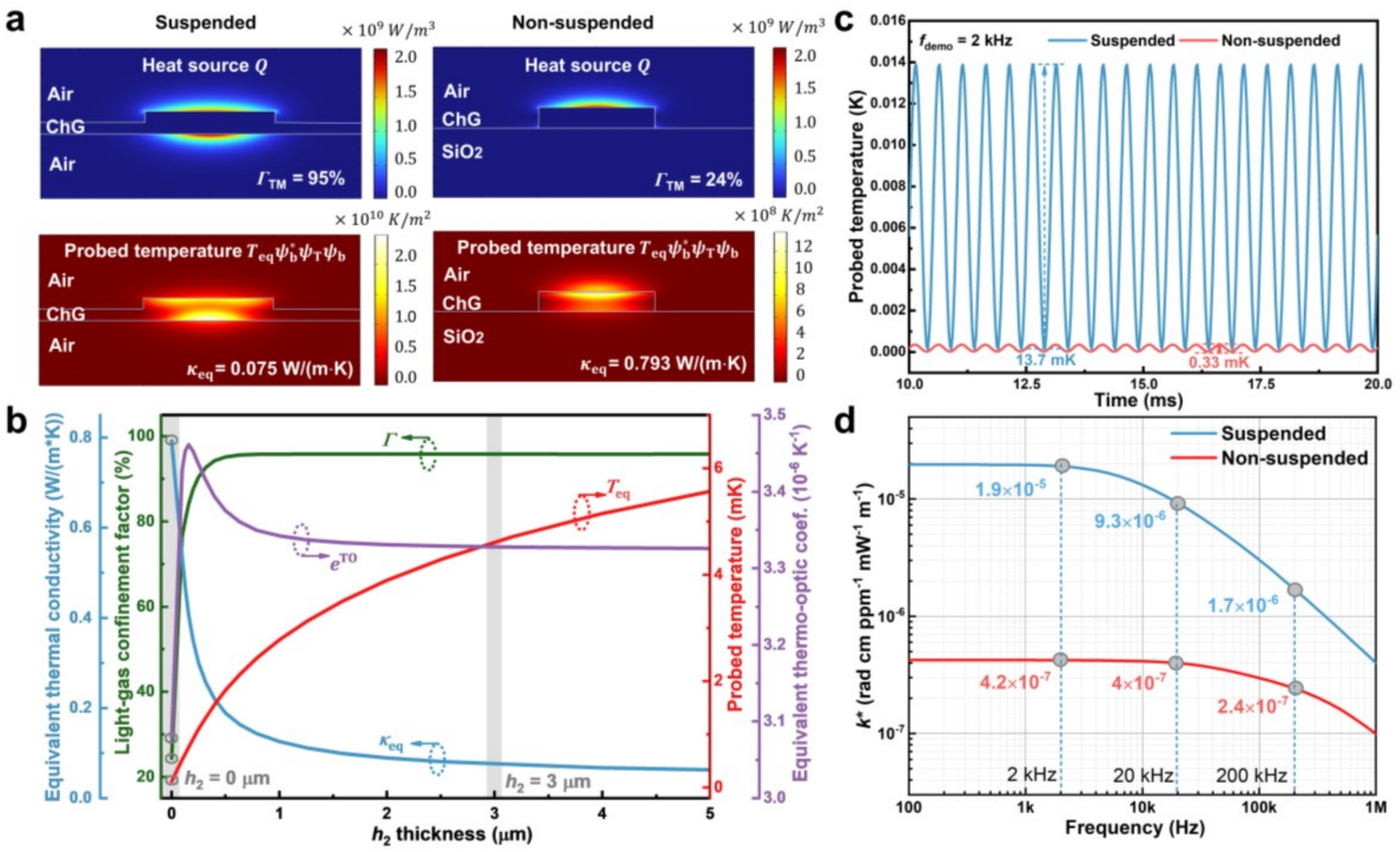

**Fig. 3 Enhancement factor evaluation. (a)** Comparison of optical (upper panel) and thermal (lower panel) confinements of the suspended and non-suspended waveguide structures. The $\Gamma$ value of the suspended waveguide is ~4 times larger than that of the non-suspended one, and $\kappa_{\mathrm{eq}}$ is ~10.6 times smaller than the non-suspended one. **(b)** Dependence of the $\Gamma$, $e_{\mathrm{eq}}^{\mathrm{TO}}$, $\kappa_{\mathrm{eq}}$ and $T_{\mathrm{eq}}$ on the thickness of the air buffer zone $h_2$. The circled points at $h_2 = 0$ μm correspond to the non-suspended structure. **(c)** Time-domain simulation at demodulation frequency of 2 kHz. The temperature variation is ~42 times larger than the non-suspended one. **(d)** Frequency-domain simulation. At 2 kHz, $k^*$ value is nearly 45 times larger than the traditional non-suspended one.

## 3. Results

**3.1 The suspended ChGW.** The suspended ChGW is fabricated following a CMOS-compatible process on a 6-inch silicon wafer with a 3-μm-thick oxide cladding.[32] A novel two-step patterning is utilized in this work to fabricate the suspended ChGW with low transmission loss and high mechanical stability. In the first patterning, holes are dry-etched on the evaporated thin film and are also used for subsequent wet etching. Ridge waveguide structures with optimized parameters are then formed using the second patterning. The process is completed by removing the underlying silicon oxide layer with hydrofluoric acid to form a suspended ChGW. Detailed processes refer to Methods and Supplementary Note 5.

The scanning electron microscope (SEM) images of the fabricated suspended ChGW are shown in Figs. 4(b)-(d). The top view of the square openings in Fig. 4(b) depicts 3×3 μm$^2$ channels used for membrane release, which also enables sufficient gas flow for light-gas interaction along the waveguide. The transmission loss of the ChGW is evaluated to be ~2.6 dB/cm (Supplementary Note 3).[33] Figs. 4(c) and (d) show the SEM cross-section views of the suspended ChGW with a rib height of ~300 nm and a width of ~1000 nm. After fabricating the suspended ChGW, a tapered silica fiber (for delivering the pump and probe beams into the ChGW) is end-faced coupled into one end of the ChGW. As shown in Fig. 4(a), the inherent reflections at the waveguide/air facets form an F-P cavity, which is used as an optical interferometer to convert the PT phase modulation into detectable intensity variations (Supplementary Note 2).[14,34]

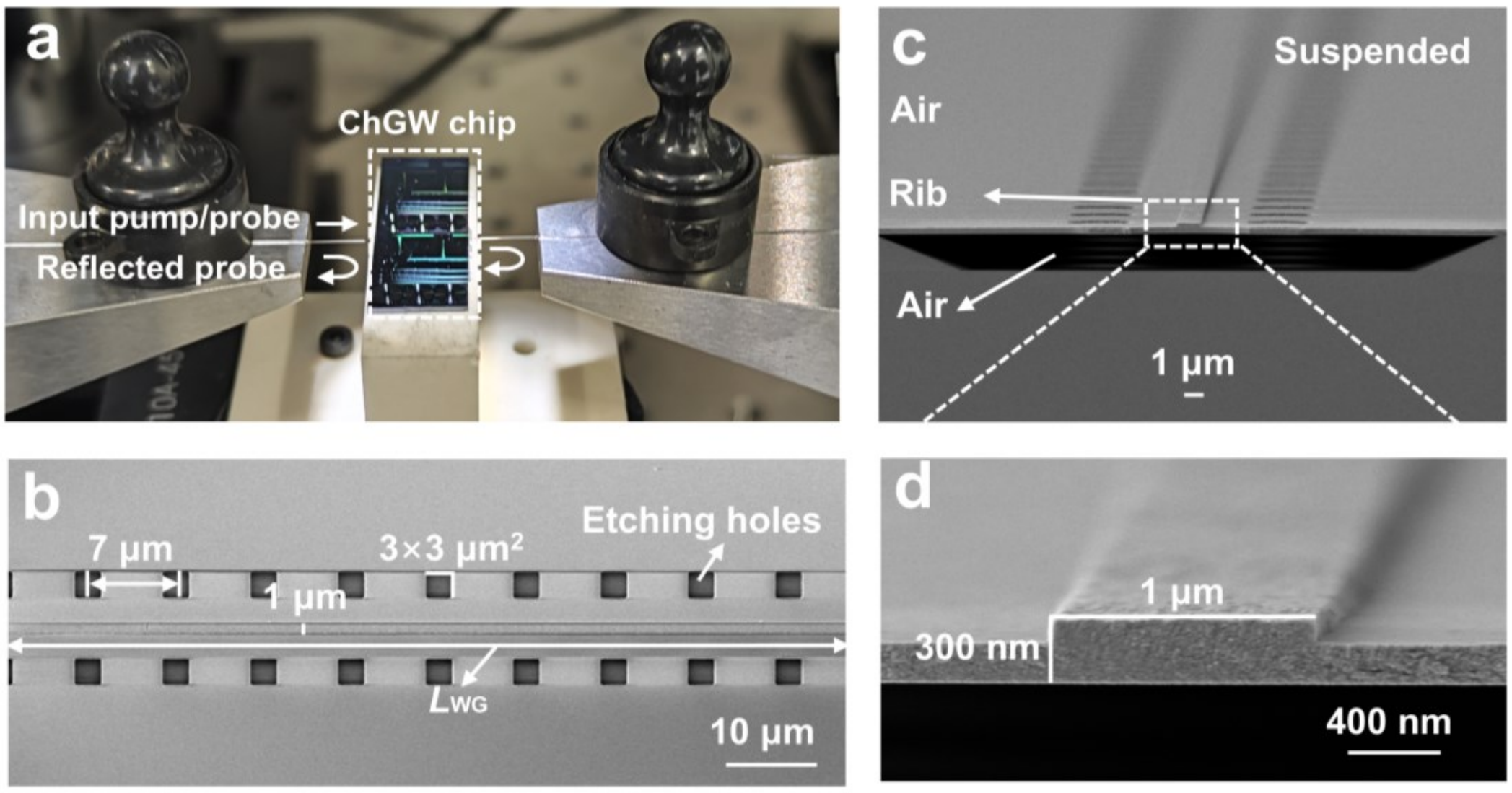

**Fig. 4 Suspended ChGW fabrication. (a)** End-face coupling platform between the tapered fiber and the suspended ChGW chip with a coupling loss of ~8 dB/facet. Such a loss is mainly due to the mismatch of the optical mode overlaps and can be optimized by exploiting inverse-tapered coupling. SEM images of the suspended chip from **(b)** top view with 3×3 μm$^2$ channels for membrane release. **(c)** Cross-section view and **(d)** Zoom-in view.

**3.2 Characterization of on-chip F-P cavity.** As depicted in Fig. 5(a), the reflection spectrum of the F-P cavity is measured using a broadband amplified spontaneous emission (ASE) source and an optical spectrum analyzer (OSA) with a wavelength resolution of 10 pm. The free spectral range (FSR) and fringe contrast of the interference spectrum around 1542 nm are approximately ~60 pm and ~8 dB, respectively. The FSR is determined using $\lambda^2/2n_mL,$ where $n_\mathrm{m}$ is the effective mode RI of the probe beam (at ~1542 nm), and $L$ is the cavity length (~1.2 cm). The resultant $n_\mathrm{m}$ is approximately 1.6, which is close to the simulated $TM_0$ mode ($n$=1.4) than the $TE_0$ mode ($n$=2.2), verifying that a $TM_0$-dominated mode is transmitted along the suspended ChGW. Fast Fourier transform (FFT) of the measured reflection spectrum reveals several minor harmonic peaks, as shown in Fig. 5(b), indicating some residual higher-order reflections may exist in the F-P cavity. More details refer to Supplementary Note 7.

Fig. 5(c) presents the servo-locking process of the F-P cavity around the quadrature point to stabilize the interferometer using a fiber laser as probe source (Details in Section 3.3). The servo amplitude stabilizes immediately after the servo stabilization is applied, and the standard deviation of the phase fluctuation is measured to be ~0.66°. We then examine the F-P cavity noise with results depicted in Fig. 5(d). The noise level is characterized by measuring probe power spectral density (PSD) on the photodetector (PD) output using an electrical spectrum analyzer (ESA). For the first case with probe beam directly into the PD without the F-P cavity (blue line), PSD reveals the relative intensity noise (RIN) of the probe laser. We then measured the PSD after the probe beam reflected from the F-P cavity (red line), where the power level of PD input is the same as in the first case. The fluctuations caused by environmental perturbations (e.g., temperature drift, mechanical instability) are typically low-frequency (below 300 Hz). We employed a servo-loop system to compensate for these low-frequency noises, ensuring the F-P interferometer remains stable at the quadrature working point. For modulation frequencies above 1 kHz, the noise level is small, and thus a pump modulation frequency of 1 kHz (with demodulation at 2 kHz) is used in the following experiments. Notably, the PSD exhibits a twofold increase at modulation frequencies above 1 kHz compared to the direct PD measurement. This difference is mainly attributed to the phase noise from the fiber laser with a 2.4-cm-long optical path difference.[35]

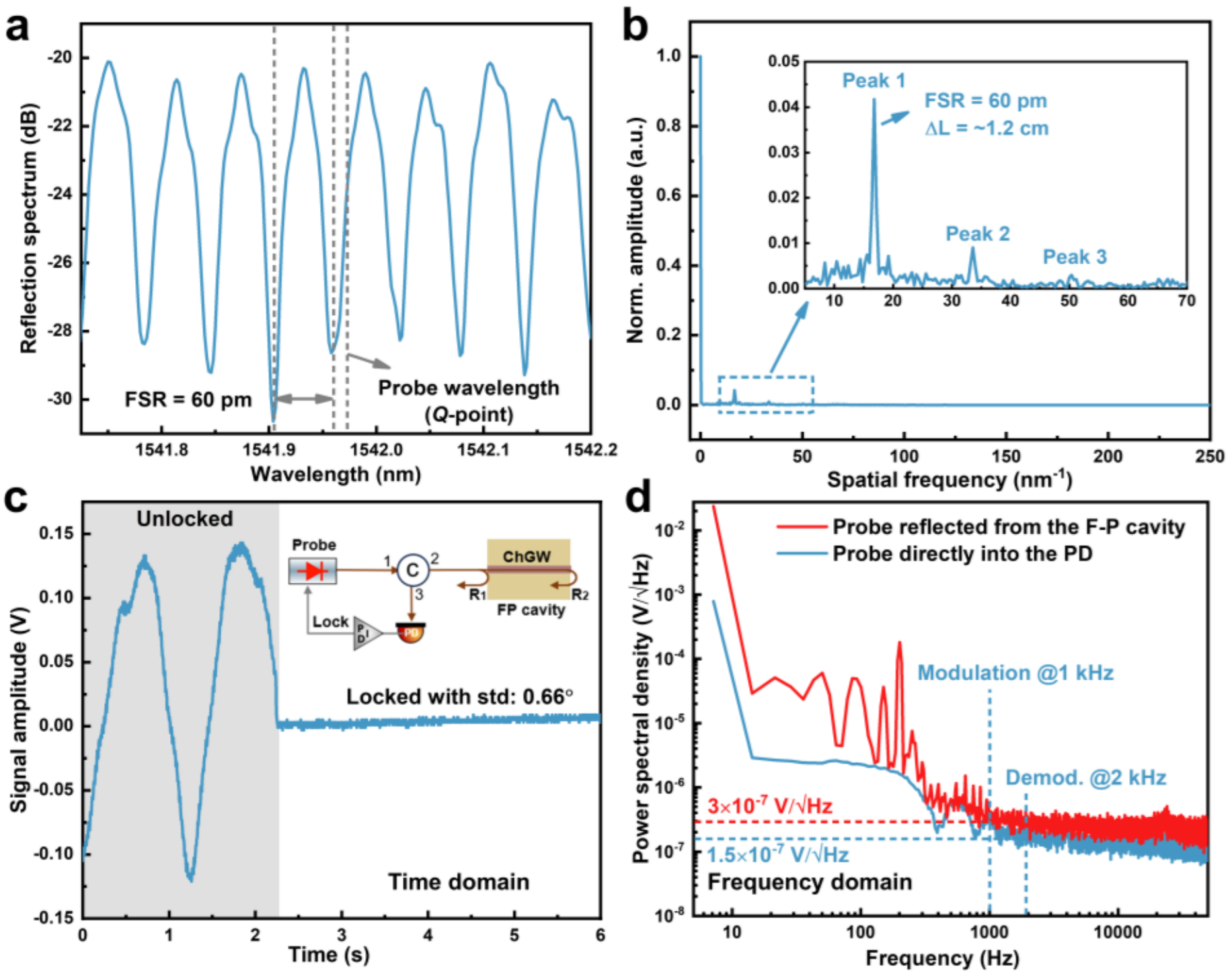

**Fig. 5 F-P cavity characterization. (a)** Reflection spectrum of the F-P cavity and **(b)** its FFT result. Inset shows three harmonic peaks caused by multiple reflections within the interface of waveguide/air, confirming the existence of F-P cavity. **(c)** Servo-locking process of the F-P cavity at the quadrature point to stabilize the interferometer. Inset shows the schematic of the F-P cavity of the probe beam. Detailed set-up is shown in Fig. 6(a). **(d)** F-P cavity noise assessment using the PSD measurement with an ESA, showing low-noise performance over 1 kHz.

**3.3 On-chip gas sensing.** The gas sensing performance of the SWE-PTS sensor is evaluated with the experimental setup shown in Fig. 6(a). Both pump and probe beams are adjusted to $TM_0$ mode using two polarization controllers (PCs). The wavelength of the distributed feedback (DFB) pump laser is sinusoidally modulated, with the center wavelength tuned to the P(11) absorption line of $C_2H_2$ at 1531.58 nm. As indicated in Fig. 5(a), a probe beam from a narrow-linewidth (~1 kHz) fiber laser with its center wavelength tuned to the quadrature point (*Q*-point) at ~1541.955 nm is used to maximize the phase-to-intensity conversion efficiency. The reflected probe beam from the ChGW is split into two parts, one is into servo-loop with proportion-integration-differentiation (PID) and low-pass filter components for servo-locking, and the other is for phase demodulation using a lock-in amplifier (LIA) to extract the 2*f* signal. For trace gas detection, $C_2H_2$ samples with different concentration levels are prepared by mixing $C_2H_2$ with $N_2$ at different flow-rate ratios.

Fig. 6(b) shows the dependence of PTS signal, noise level and measured signal-to-noise ratio (SNR) on the pump modulation frequency. The PTS signal decreases with increasing modulation frequency and the 3-dB roll-off frequency is about ~8 kHz. The noise is measured by filling the gas cell with $N_2$, and the maximum SNR is achieved around 1 kHz, hence we select 1kHz in subsequent experiments. We validate the sensor performance by increasing the pump power inside the ChGW, as shown in Fig. 6(c). A linear relationship between PTS signal and input pump power is obtained with an $R^2$ of 0.999. To further assess the detection limit of the SWE-PTS sensor, Allan-Werle analysis is conducted based on the noise data with $N_2$ filling into the gas cell. As shown in Fig. 6(d), the minimum Allan deviation is achieved at an averaging time of 65 s, where the lowest noise level is estimated to be ~22 nV, corresponding to a detection limit of 330 ppb for $C_2H_2$ detection. This yields a noise-equivalent absorption (NEA) of $3.8\times10^{-7}$ $cm^{-1}$, and a normalized noise-equivalent absorption (NNEA) of $9.9\times10^{-9}$ $cm^{-1}\cdot W\cdot Hz^{-1/2}$.

Fig. 6(e) shows the peak-to-peak values of the 2*f* signal as functions of $C_2H_2$ concentrations from 0 to 30%. The 2*f* signal increases approximately linearly with the gas concentration, giving a dynamic range of nearly 6 orders of magnitude (~$9.1\times10^5$). The response time is tested by sequentially filling the gas cell with $N_2$, 330 ppm $C_2H_2$, and then $N_2$ at a flow rate of 200 standard cubic centimeters per minute (sccm), while the pump wavelength is fixed at 1531.58 nm. Fig. 6(f) shows the real-time recording of the PTS signal. The response time, specifically the time required for the response reaching 90% (*i.e.*, rise time) and descending to 10% (*i.e.*, fall time) of the maximum PTS signal, is less than 1 s. Such a response time is primarily limited by the gas exchange time, which is largely dependent on the volume of the gas chamber (~0.5 mL). Faster response can be obtained by increasing the inlet pressure difference or using a pump at the outlet, also an on-chip microfluidic channel cell would certainly help reduce it to ~ms scale.[36,37]

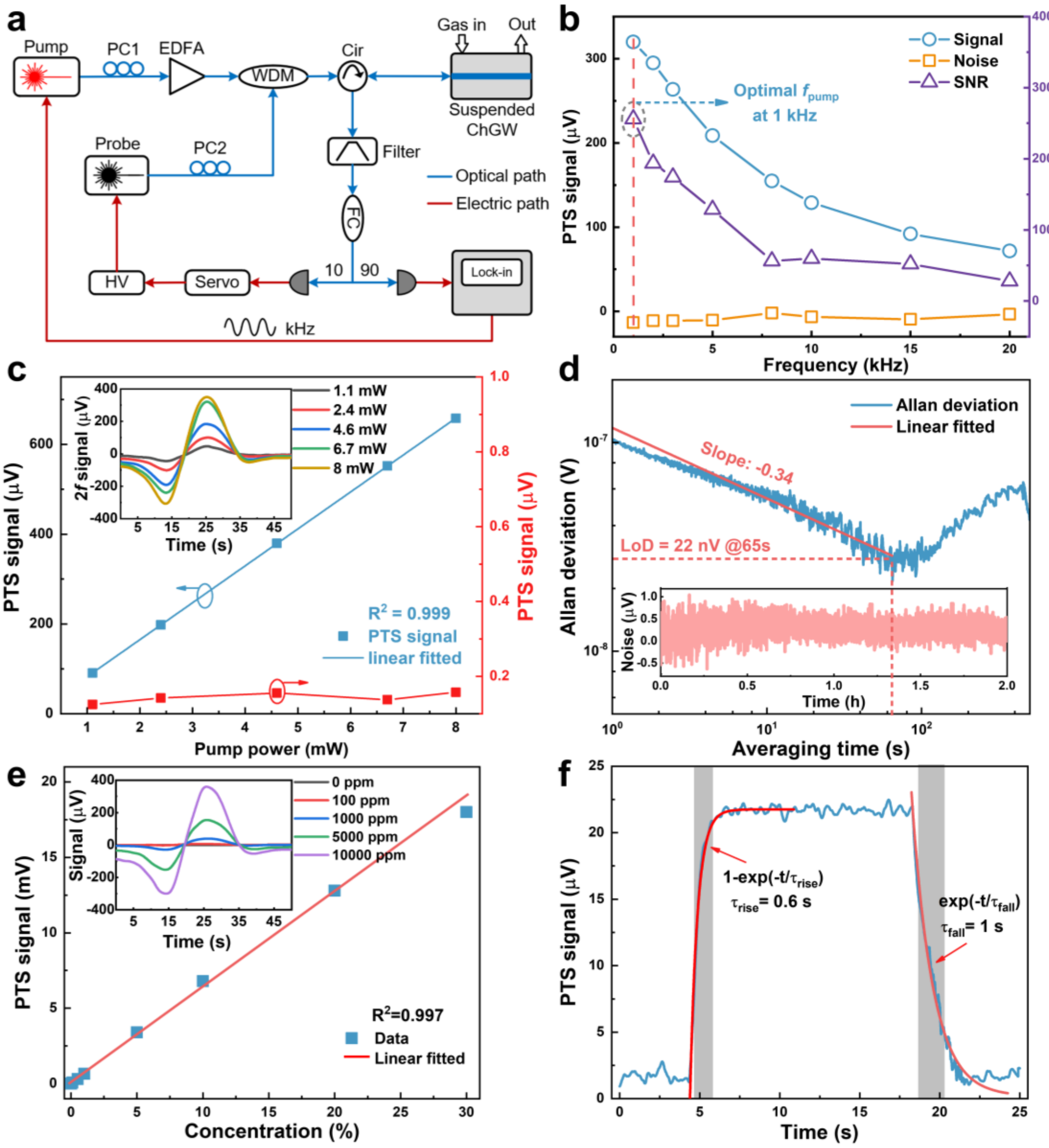

**Fig. 6 Gas sensing performance. (a)** Experimental set-up of the SWE-PTS. EDFA: erbium-doped fiber amplifier, Cir: optical circulator, FC: fiber coupler, PD: photodetector. **(b)** Measured PTS signal amplitude, standard deviation (std.) of the noise and SNR as a function of pump modulation frequencies. **(c)** Detected 2*f* signal amplitude at *f* =1 kHz and std. of the noise as functions of the pump power. Inset are the 2*f* signals when the pump wavelength is tuned across the $C_2H_2$ lines. **(d)** Allan-Werle plot based on the noise data over a period of 2 h, which is shown in the inset. For the noise measurement, the lock-in time constant is 100 ms, corresponding to 0.94 Hz detection bandwidth. **(e)** Dynamic range evaluation with peak-to-peak value of the 2*f* signal as a function of gas concentrations. Inset shows the 2*f* signals for $N_2$, 100 ppm, 1000 ppm, 0.5%, and 1% $C_2H_2$ when the pump power is set at 8 mW. The detection bandwidth is 0.094 Hz. **(f)** Response time evaluation. The PTS signal amplitude rises exponentially to a flat level following $1-\exp(-t/\tau_{rise})$ with the rise time $\tau_{rise}$, and then follows an exponential decay in the form of $\exp(-t/\tau_{fall})$ with the rise time $\tau_{fall}$. At ~4 s, 330 ppm $C_2H_2$ gas is loaded into the gas cell at a flow rate of 200 sccm, which is prefilled with $N_2$. At ~19 s, $N_2$ is refilled into the gas cell at the same flow rate.

## 4. Discussion

In summary, we have reported a novel suspended ChGW nanophotonic platform for highly sensitive on-chip PTS gas molecular sensing. Theorical and experimental results verified its superiority by improving the efficiency of heat generation and heat accumulation. Guided by an equivalent PT phase modulation model, we developed an optimized suspended ChGW that exhibits a nearly 4 times higher heat source power and a 10.6 times lower equivalent heat conductivity as compared to conventional strip waveguides. These factors work together to achieve 42 times higher probed temperature variation, leading to 45-fold larger PT phase modulation. We successfully demonstrated high sensitivity $C_2H_2$ sensing with a detection limit of 330 ppb, a dynamic range spanning 6 orders of magnitude, and a fast response of less than 1 s. These comprehensive performance advantages pave the way for high-sensitivity, high-selectivity, wide dynamic range, rapid response, and ultracompact on-chip gas sensors, representing a significant step forward in the development of fully integrated photonic sensors.

To benchmark the performance metrics of the reported SWE-PTS, we compare it with the representative state-of-the-art waveguide sensors, as shown in Fig. 7.[10,11,13,19-21,24,38-44] Detailed comparisons are provided in Supplementary Note 9. These sensors cover different analyte gases, system configurations and parameters. Here, we focus on comparing the sensor performance in terms of NEA and dynamic range. With a 1.2-cm-long suspended waveguide, the SWE-PTS sensor achieves NEA of $3.8\times10^{-7}$ $cm^{-1}$, representing improvement of 1−4 orders of magnitude over existing waveguide sensors. The dynamic range reaches nearly 6 orders of magnitude, surpassing prior on-chip gas sensing technologies by more than 2 orders of magnitude. These remarkable results indicate that our sensor is the first photonic waveguide sensors capable of achieving ppb-level sensitivity in the NIR range. This is particularly significant given the orders of magnitude weaker absorption characteristics of the NIR compared to the MIR, as well as the relatively short interaction length of only 1.2 cm on a nanophotonic chip.

Better performance could be realized by further optimizing the sensor parameters. The transmission loss of ~2.6 dB/cm in this suspended ChGW is primarily resulted from scattering due to the waveguide sidewall roughness. To minimize surface roughness, a waveguide thermal reflow process prior to suspension has been proved as an effective approach.[45] It is with great potential the waveguide loss can be reduced down to ~1 dB/cm, which potentially translate into an improved detection limit of less than 100 ppb. Additionally, ChG possess an ultrawide transmission range from 1 to 16 μm, which would allow the detection of multiple molecular species in the NIR and MIR, including greenhouse gases and volatile organic compounds (Supplementary Note 8).[46] Such a SWE-PTS system is readily to be extended into MIR range with 2−3 orders of magnitude performance improvement using the strongest molecular absorption. Moreover, the extension of on-chip SWE-PTS to solid and liquid analysis is straightforward. Planar suspended platforms enable the investigation of photothermal, optomechanical, and optoacoustic effect in solid and liquid medium, thereby significantly advancing the fields of interface optics[1] and chemical reaction dynamics.[2] We also foresee the developed on-chip SWE-PTS as a highly potent technology for environmental monitoring,[47] health screening,[48] and wearable device applications.[49]

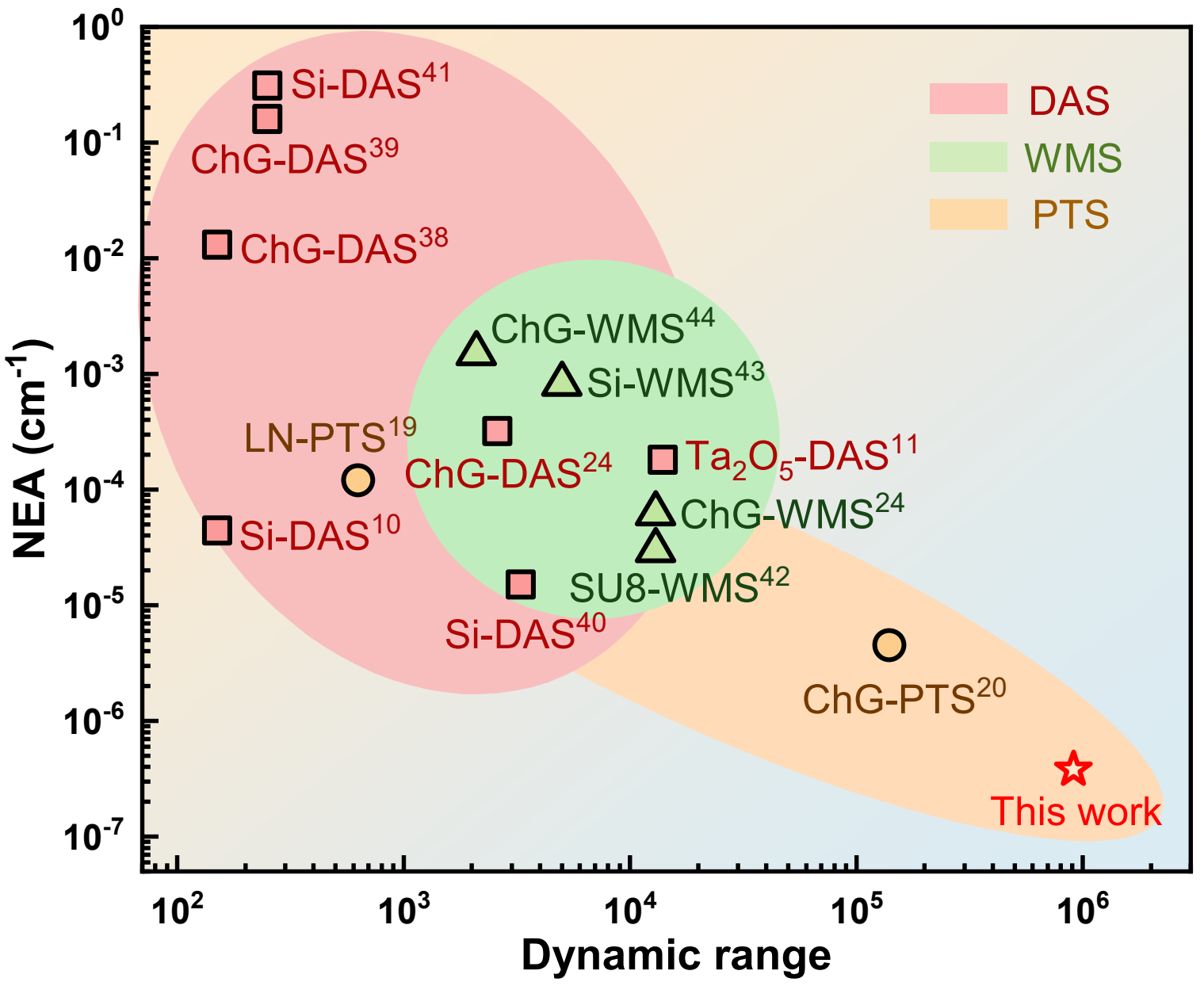

**Fig. 7 Performance comparison.** Comparison of the SWE-PTS sensor with the state-of-the-art waveguide sensors in terms of NEA and dynamic range. DAS: direct absorption spectroscopy. WMS: wavelength modulation spectroscopy, a variation of DAS. Si: silicon, SU8: polymer photoresist, $Ta_2O_5$: tantalum pentoxide, LN: lithium niobate.

## Methods

**Suspended ChGW fabrication.** A 6-inch CMOS-compatible fabrication line was used for suspended ChGW with a two-step patterning process. In the first step, the sacrificial silicon dioxide and ChG film was deposited on to the thermally oxidized silicon wafer by chemical vapor deposition and thermal evaporation, respectively. In addition, arrays of micro-holes were patterned and dry-etched by inductive-coupled plasma (ICP) etcher for the final membrane release. The resist is then removed by soaking into (1-methyl-pyrodine) NMP and isopropanol (IPA) with the aid of ultrasonication. The second step involves the precise registration of the ChG rib waveguide in between the micro-hole array. The rib waveguides were subsequent etched in ICP and the corresponding photoresist is stripped. The final step involves immersing the waveguide in 5% HF, undercutting the thermally oxidized silicon through the micro-holes prepared in the first layer. This etching process is meticulously timed and controlled to prevent over-etching and structural damage. More details refer to Supplementary Notes 5 and 6.

**Fabrication tolerance and scalability estimation.** To evaluate the fabrication tolerance, the cross-section metrology shows deviations of approximately 20–30 nm in waveguide width *w*, and around 3 nm in core-layer height *h*. The resulting fractional change in and hence PTS SNR can be estimated with the aid of Fig. 2, where the overall performance variation is typically <1%. Among these geometric parameters, the core-layer height is the most sensitive and therefore requires more precise control during fabrication. The scaling to multi-species gas detection sensor chip can be made with waveguide arrays aiming at various detection wavelengths. Taking advantage of the CMOS technology, reliable, wafer scale production is possible. The biggest challenge is the undercladding BOX removal process. It requires dipping the wafer into HF solution for precise time control. In addition, rinsing and cleaning may include the use of critical point drier that helps retaining large area suspended membrane without suffering through the high surface tension of water during drying.

**Gas cell and gas delivery system.** The ChGW is sealed by a 3D-printed gas cell with two ports for gas exchange. One face is milled to form a downward-facing chamber and then inverted to cap the waveguide chip, defining a dimension of $1\times1\times0.5$ $cm^3$. Sample gas is delivered via 2 mm inner-diameter clear silicone tubing to the gas cell. Upstream of the cell, a 1.2 m straight run is included to ensure thermal equilibration and eliminate Joule-Thomson cooling effects. Using a simple convective heat-transfer estimate, for a pressure drop ~2 bar and a flow rate of 200 sccm, the characteristic thermal equilibration length is only ~3.5 cm and the Peclet number $Pe \sim 10^{-2} \ll 1$ , so the heat convection can be ignored and the gas region can be considered at ambient temperature upon entering the cell. Inside the gas cell, the ChG waveguide is placed at the bottom of the gas chamber, approximately 3.5 mm below the jet axis of the gas inlet. The inlet port is positioned at the geometric center of the ceiling to avoid a focused jet directly impinging on the chip surface. Since the waveguide is not directly exposed to the gas jet, it produces only a gentle background velocity near the chip. The additional forced-convection heat loss is therefore negligible compared with conduction-dominated heat transfer.

**Numerical simulation.** The photothermal transient time-domain profile and steady state frequency-domain simulation are investigated using the finite element method (FEM) with COMSOL Multiphysics. The gas flow in the evanescent region is considered as a continuum, following the physical laws of thermodynamics, diffusion, and thermal conductivity. Since the gas flow rate is relatively slow, thermal conduction is considered as the main dissipation mechanism in this PT model, where the convection and radiation are neglected. We use the thermal conductivity of air for the gas medium, the value of which is 0.026 W/(m·K) at ambient temperature and is temperature-dependent. The limit that can optimize thermal conductivity through suspended design is that both the $SiO_2$ and Si layer are completed removed (etched away, which is practically impossible) and the corresponding value of $\kappa_{\mathrm{eq}}$ is then 0.0344 W/(m·K), only about 2.2 times larger than the actual suspended ChGW in this case. Another consideration is that the trace gas absorption is regarded as weak ($\alpha\Gamma CL$<<1) so that the pump depletion along the waveguide can be safely ignored. For our case here, under conditions of $\alpha$=1.05 $cm^{-1}$, $\Gamma$=95%, $C$=1% and $L$=1.2 cm, the relative error is <1%, which satisfies the weak absorption condition.

**Acknowledgements**

The authors wish to express their gratitude to the support of Hong Kong SAR government GRF grant (Grant No. 15223421), the Photonics Research Institute of The Hong Kong Polytechnic University (PolyU) (Grant No. 1-CDJ6), the National Natural Science Foundation of China (Grant Nos. 62305304, 62005233, 62175087, 62235016, 62535009), the National Key R&D program (Grant No. 2021ZD0109904), the RGC-CRF funding (Grant No. C4002-22Y), the Natural Science Foundation of Jilin Province (Grant No. YDZJ202501ZYTS295). The authors also acknowledge the Nano Fabrication Center of Zhejiang Laboratory for device fabrication and characterization.

## Author contributions

K.Z. conceived the idea. H.L. developed the theory. K.Z. together with H.L. performed the mathematical formulations and experiments. F.H., Y.Z. and J.G. fabricated the suspended waveguide and performed morphological characterization. X.W. P.Z., and H.B. assist in numerical simulation. S.Y., L.L. and L.W. analyzed the data. K.Z. and Q.D. wrote the original manuscript. Q.D. C.Z. and W.J. supervised the work. All authors reviewed the manuscript and provided editorial input. K.Z., H.L. and F.H. contributed equally to this work.

## Data availability

The data that support the plots within this paper and other finding of this study are available from the corresponding authors upon reasonable request.

## Conflict of interest

The authors declare that they have no conflict of interest.

**Supplementary information is available for this paper.**

# Supplementary information for
# Suspended waveguide-enhanced near-infrared photothermal spectroscopy for ppb-level molecular gas sensing on a chalcogenide chip

Kaiyuan Zheng[1,2,†], Hanyu Liao[1,2,†], Fengbo Han[3,†], Xueying Wang[4], Yan Zhang[3], Jiaxin Gu[3], Pengcheng Zhao[1,2], Haihong Bao[1,2], Shaoliang Yu[3], Qingyang Du[3*], Lei Liang[5], Chuantao Zheng[4*], Wei Jin[1,2*], and Lijun Wang[5]

[1]Department of Electrical and Electronic Engineering and Photonics Research Institute, The Hong Kong Polytechnic University, Hong Kong, 999077, China

[2]Photonics Research Center, The Hong Kong Polytechnic University Shenzhen Research Institute, Shenzhen, 518071, China

[3]Zhejiang Lab, Hangzhou, 311121, China

[4]State Key Laboratory of Integrated Optoelectronics, JLU Region, College of Electronic Science and Engineering, Jilin University, Changchun, 130012, China

[5]State Key Laboratory of Luminescence and Applications, Changchun Institute of Optics, Fine Mechanics and Physics, Chinese Academy of Sciences, Changchun 130033, China

[†]These authors contributed equally to this work

[*]Corresponding authors: qydu@zhejianglab.edu.cn, zhengchuantao@jlu.edu.cn, wei.jin@polyu.edu.hk

## Note 1. Theory of the suspended waveguide-enhanced PTS (SWE-PTS)

The suspended waveguide enhances the performance of the photothermal spectroscopy (PTS) by improving the efficiency of two physical processes: the enhanced pump absorption that produces a more efficient heat source, and the enhanced heat accumulation near the waveguide region that results in higher PT phase modulation efficiency. These enhancements depend on the dimensions of the suspended waveguide as well as the pump and probe wavelengths. For simplicity, we ignore the surface roughness and assume waveguide is uniform along the *z*-axis.

(i) We first discuss the enhancement of pump absorbance that produces heat source. In general, for a monochromatic pump with its wavelength centered at the gas absorption line in free-space, the heat source is a time-invariance source that is equal to $\alpha C I_{\mathrm{p}}$, $\alpha$ is the gas absorption coefficient, $C$ is the gas concentration, and $I_{\mathrm{p}}(x,y)$ is the pump intensity. For the heat source in SWE-PTS, the pump is wavelength-modulated around the gas absorption line, and the light-gas interaction is actuated by the pump evanescent field in gas. Hence, the corresponding heat source is time-periodic and distributed outside the waveguide material. Similar to the wavelength modulation spectroscopy (WMS),[1], the heat source can be decoupled as a sum of $n^{\mathrm{th}}$ order harmonic sources as:

$$Q(x,y,t)=\sum_{n} Q_{\mathrm{n}}=\sum_{n} H_{\mathrm{n}} \alpha C I_{\mathrm{p}}^{\mathrm{gas}}(x,y) e^{\mathrm{jn\omega t}} \tag{S1}$$

here $H_n$ is the harmonic modulation coefficient, $\alpha$ is the peak absorption coefficient of the selected gas absorption line, $C$ the concentration of the gas, $\omega$ the modulation frequency of the pump beam. The $I_{\mathrm{p}}^{\mathrm{gas}}$ used here is the intensity of the pump beam with the superscript denotes the gas region, which is defined through:

$$I_{\mathrm{p}}^{\mathrm{gas}}(x,y)=\frac{1}{2}c\varepsilon_0 n_{\mathrm{gas}}\left|E_p(x,y)\right|^2 \quad \text{(S2)}$$

here $c$ is the speed of light, $\varepsilon_0$ is the dielectric constant, $n_{\mathrm{gas}}$ is the refractive index of the gas material, $E_{\mathrm{p}}(x,y)$ is the pump electric field. For gas absorption line in the profile of Lorentz function, the WMS technique picks the second order harmonic signal $n=2$ with $H_2=0.343$ at optimized modulation parameters. Hence, we focus on the second harmonic heat source $Q_2$. To compare the heat generation efficiency among different waveguide structures, we introduce the heat source power $P_{\mathrm{Q}}$, which is defined as:

$$P_{\mathrm{Q}}=\iint_{\mathrm{total}} Q_2\, dxdy \quad \text{(S3)}$$

here $P_{\mathrm{Q}}$ is in the unit of Watt that denotes the energy transferred from pump to thermal energy. Combine Eqs. (S1-S3), the heat source power is related to the intensity of the pump evanescent field as:

$$P_{\mathrm{Q}}=H_2\alpha C e^{\mathrm{j}2\omega\mathrm{t}}\iint_{\mathrm{gas}} I_{\mathrm{p}}^{\mathrm{gas}}\, dxdy \quad \text{(S4)}$$

Since $I_{\mathrm{p}}^{\mathrm{gas}}$ is only a fraction of the total pump intensity, it is convenient to express the integration of $I_{\mathrm{p}}^{\mathrm{gas}}$ by using the measurable pump power $P_{\mathrm{Q}}$ through:

$$\iint_{\mathrm{gas}} I_{\mathrm{p}}^{\mathrm{gas}}\, dxdy = P_p\Gamma \quad \text{(S5)}$$

Here the variable $\Gamma$ is some fraction that scales the heat source power to the pump power, which is exactly the gas confinement factor (GCF)[2]. The definition of $\Gamma$ immediately follows Eq. (S5) as:

$$\Gamma := \frac{\iint_{\mathrm{gas}} I_{\mathrm{p}}^{\mathrm{gas}}\, dxdy}{P_{\mathrm{p}}} = \frac{n_{\mathrm{gp}}\iint_{\mathrm{gas}} \varepsilon(x,y)\left|E_{\mathrm{p}}(x,y)\right|^2 dxdy}{n_{\mathrm{gas}}\iint_{\mathrm{total}} \varepsilon(x,y)\left|E_{\mathrm{p}}(x,y)\right|^2 dxdy} \quad \text{(S6)}$$

here $n_{\mathrm{gas}}$ is refractive index (RI) of gas, and we have used the definition of $n_{\mathrm{gp}}=cU_{\mathrm{p}}/P_{\mathrm{p}}$ in deducting Eq. (S6), where $U_{\mathrm{p}}$ is the pump energy stored in the fiber cross-section. Based on Eq. (S5) and Eq. (S6), one can define the normalized pump field distribution $\psi_{\mathrm{p}}$ as:

$$\psi_{\mathrm{p}}=\frac{\sqrt{\varepsilon_0}n(x,y)E_{\mathrm{p}}(x,y)}{\iint_{\mathrm{total}} \varepsilon_0 n(x,y)^2\left|E_{\mathrm{p}}(x,y)\right|^2 dxdy} \quad \text{(S7)}$$

$$\Gamma=\frac{n_{\mathrm{gp}}}{n_{\mathrm{gas}}}\iint_{\mathrm{gas}} \psi_{\mathrm{p}}^*\psi_{\mathrm{p}}\, dxdy=\frac{n_{\mathrm{gp}}}{n_{\mathrm{gas}}}\gamma_{\mathrm{p}} \quad \text{(S8)}$$

here $\psi_{\mathrm{p}}$ is simply the electric field $E_{\mathrm{p}}$ normalized by its stored energy $U_p$, and we have also introduced the evanescent field ratio $\gamma_{\mathrm{p}}$ representing the fraction of the pump intensity in the gas region:

$$\gamma_{\mathrm{p}} := \frac{\iint_{\mathrm{gas}} \psi_{\mathrm{p}}^*\psi_{\mathrm{p}}\, dxdy}{\iint_{\mathrm{total}} \psi_{\mathrm{p}}^*\psi_{\mathrm{p}}\, dxdy} \quad \text{(S9)}$$

Combining Eq. (S4) with Eqs. (S4-S9), the power of the heat source can be expressed as:

$$P_{\mathrm{Q}}=H_2\alpha C P_{\mathrm{p}}\Gamma e^{\mathrm{j}2\omega\mathrm{t}} \propto \frac{n_{\mathrm{gp}}}{n_{\mathrm{gas}}}\alpha C P_{\mathrm{p}}\gamma_{\mathrm{p}} \quad \text{(S10)}$$

hence the suspended waveguide enhances the power of the heat source in the PTS with a higher $\Gamma$. This enhancement attributes to a larger proportion of the evanescent field in the gas region.

For the suspended waveguide used in this work, the GCF could be 3−4 times larger as compared to the non-suspended waveguide.

(ii) We discuss the enhancement of heat accumulation that improves photothermal phase modulation. The second-order harmonic heat sources $Q_2(x, y, t)$ results in a temperature change in the form of $\Delta T(x, y, t) = \tilde{T}(x, y)e^{-j2\omega t}$ through the heat conduction, which lead to subsequent probe phase modulation at the same frequency. Assuming that $\alpha CL \ll 1$ for trace gas detection, $L$ is the length of the waveguide, the heat transfer equation can be solved in the form of linear perturbation. The resulting temperature change may be determined by solving the heat transfer equation for $\tilde{T}$ in the frequency domain:

$$Cj2\omega\tilde{T} - \nabla \cdot (\kappa\nabla\tilde{T}) + C\vec{v} \cdot \nabla\tilde{T} = \tilde{Q}_2 \tag{S11}$$

here $C = \rho c$ is the volumetric heat capacity with $\rho, c$ being the gas density and heat capacity, $\kappa$ is heat conductivity, $\vec{v}$ is the gas flow velocity which is set to be zero in solid region. The ~ denotes the amplitude of the variables in the frequency domain, *i.e.* $\tilde{Q}_2 = H_2\alpha C I_{\mathrm{p}}^{\mathrm{gas}}$. To analytically solve Eq. (S11), we need to make some assumptions first. Since the non-zero gas velocity $\vec{v}$ would improve the heat conduction and reduce the heat accumulation, in sensor fabrication the ChG waveguide is placed at the bottom of the gas chamber, approximately 3.5 mm below the jet axis of the gas inlet. By doing so, the local gas velocity near the waveguide is governed by the net through-flow rather than by the nozzle exit velocity, which is typically small and can be safely ignored. In our experiment, we employed 200 sccm inlet and outlet in a 1 cm×1 cm×0.5 cm gas cell. The cross-sectional area is $A_{cell} = 5 \times 10^{-5}\, m^2$ and $F_{out} = 200\, sccm$. Away from the jet core, the background "sweep" velocity can be estimated by:

$$v \approx \frac{F_{out}}{A_{cell}} \approx 6.7\, cm/s \tag{S12}$$

Whether convection significantly alters the heat dissipation is determined by the Péclet number:

$$Pe = \frac{vL_e\rho c}{\kappa} \tag{S13}$$

with $L_e$ being the relevant thermal length scale. For heat transfer equation solved at frequency domain, the appropriate length scale is the thermal penetration depth as:

$$L_e = \delta_i = \sqrt{\frac{2\kappa_i}{\rho_i c_i \omega_m}} \tag{S14}$$

At frequency $f = 2$ kHz, the calculated Péclet number for air is $Pe \sim 10^{-2} \ll 1$, indicating that conduction dominates and convection has a negligible effect under our experimental conditions. Hence, the combination of off-axis geometry, small local velocities, and small thermal penetration depth at modulation frequencies of interest ensures that heat conduction is the governing mechanism. Besides, we neglected the interfacial thermal resistance (Kapitza resistance $R_{\mathrm{K}}$) which is typically in the range of $10^{-5} - 10^{-6}$ m$^2$K/W, much smaller than the thermal resistance of ChG ($3.8 \times 10^{-6}$ m$^2$K/W) and gas ($1.0 \times 10^{-3}$ m$^2$K/W).

To get the analytic solution of $\tilde{T}$, we introduce an equivalent heat transfer model based on Eq. (S11). For a fixed modulation frequency and fixed pump wavelength (fixed $\psi_{\mathrm{p}}$), increasing the power $P_{\mathrm{Q}}$ does not affect the temperature field profile but linearly increases the temperature change amplitude at any arbitrary point. To show this, we decompose $\tilde{T}(x, y)$ into a product:

$$\tilde{T}(x, y) = T_{\mathrm{eq}}\psi_T(x, y) \tag{S15}$$

here $\psi_T(x, y)$ is spatial profile that captures the distribution's geometric dependence, and $T_{eq}$ is the scalable equivalent amplitude. The decomposition is valid since the fixed modulation frequency and pump wavelength define a specific eigen mode $\psi_T$ for the heat transfer system. The integration form of Eq. (S11) relates the temperature change $\tilde{T}$ to the heat source power defined by Eq. (S3) as:

$$-j2\omega_m T_{eq} \iint_{total} C\psi_T dxdy - T_{eq} \iint_{total} \nabla \cdot (\kappa \nabla \psi_T) dxdy = \iint_{total} \tilde{Q}_2\, dxdy = P_Q \quad \text{(S16)}$$

where we have substituted Eq. (S15) and neglected the heat convection term. In general, direct spatial integration of Eq. (S11) to derive Eq. (S16) imposes a fundamental constraint of global thermal energy balance. The operation is generally invalid without explicitly verifying boundary compatibility, since it requires the net heat flux through boundaries to exactly offset the integration source term. In SWE-PTS, however, such a balance is physically achievable as the boundaries permit heat exchange, e.g. under constant-temperature boundary conditions or in unbounded domains where the temperature field decays at infinity. In our model, we assume the boundary at constant temperature 298 K, hence Eq. (S16) is appliable. This assumption might be invalid under the case where the photonic chip could exchange heat with the surrounding gas via natural convection, or the inlet gas flow introduces forced advection. Whether the heat convection dominates the heat transfer process can be estimated by the Peclet number, which in our case (1 cm×1 cm×0.5 cm gas cell and 200 sccm gas inlet) is $P_e \sim 10^{-2} \ll 1$, hence the heat conduction dominates and convection is negligible.

The equivalent heat transfer model Eqs. (S15-S16) is obtained from spatial coordinate decomposition and direct integration. However, the $\psi_T$ here is a general temperature profile whose normalization condition has not been specified. For SWE-PTS, we are only interested in the overlap between the probe field $\psi_b$ and the temperature field. Thus, similar to the coupled-mode theory, we choose to normalize $\psi_T$ by projecting it onto the probe mode under a weighted inner product:

$$\iint_{total} \psi_b^* \psi_T \psi_b\, dxdy \coloneqq 1 \quad \text{(S17)}$$

This normalization ensures unit coupling strength between $\psi_T$ and the probe mode $\psi_b$ in SWE-PTS, i.e., the efficiency of temperature field $\psi_T$ coupled to phase modulation of $\psi_b$ is unit. It is now sufficient to deduce the mathematical form of $T_{eq}$ by multiplying $T_{eq}$ on both sides of Eq. (S17):

$$T_{eq} = \iint_{total} \psi_b^* \tilde{T} \psi_b\, dxdy \quad \text{(S18)}$$

Eq. (S18) demonstrates the physical meaning of $T_{eq}$ under our normalization condition Eq. (S17): $T_{eq}$ is the equivalent temperature amplitude that is probed by the probe beam. From now on, $T_{eq}$ is referred as probed temperature. Combining Eq. (S16) with Eq. (S18), it is also possible to define the equivalent volumetric heat capacity $C_{eq}$ and equivalent heat conductivity $\kappa_{eq}$ that converts the heat source energy to the probed temperature:

$$C_{eq} \coloneqq \iint_{total} C\psi_T dxdy = \iint_{total} \rho c \psi_T dxdy \quad (S19)$$

$$\kappa_{eq} \coloneqq \iint_{total} \nabla \cdot (\kappa \nabla \psi_T) dxdy \quad \text{(S20)}$$

As shown in Eqs. (S19-S20), the defined $C_{eq}$ and t$\kappa_{eq}$ are exactly the volumetric heat capacity and heat conductivity experienced by the temperature field mode field $\psi_T$. As a result, the solution of the heat transfer equation is:

$$T_{\mathrm{eq}} = \frac{P_Q}{j2\omega_m C_{eq} - \kappa_{eq}} \tag{S21}$$

Eq. (S19) is the general solution under demodulation frequency of $2\omega_m$. To achieve maximum photothermal phase modulation efficiency, we choose the demodulation frequency $\omega_m = 1kHz$ which is much smaller than the 3 dB roll-off frequency. For this case, we have $j2\omega_m C_{eq} \ll \kappa_{eq}$. In this case, the thermal inertia term can be neglected and Eq. (S21) becomes a compact form:

$$T_{\mathrm{eq}} \approx -\frac{P_Q}{\kappa_{eq}} \tag{S22}$$

Numerical calculation reveals that the error of using Eq. (S22) to replace Eq. (S21) at 1 kHz is only 0.5%. According to Eq. (S20), $\kappa_{\mathrm{eq}}$captures the whole thermal conduction characteristics on the waveguide cross-section plane $dxdy$. For example, if the waveguide material is uniform and has a heat conductivity of $\kappa_0$, for a point source $Q_2$ with a low modulation frequency $\omega_m \ll \kappa_{eq}/2C_{eq}$ , Eqs. (S16, S20) yields:

$$\iint_{\mathrm{total}} \nabla^2 \psi_{\mathrm{T}}\, dxdy = 1 \tag{S23}$$

$$\kappa_{\mathrm{eq}} = \iint_{\mathrm{total}} \nabla \cdot (\kappa_0 \nabla \psi_{\mathrm{T}}) dxdy = \kappa_0 \tag{S24}$$

where the equivalent heat conductivity $\kappa_{\mathrm{eq}}$ is exactly the material conductivity $\kappa_0$. Based on our equivalent heat transfer model, the probed temperature $T_{\mathrm{eq}}$ (equivalent temperature amplitude) can be expressed based on Eq. (S10) and (S22):

$$T_{\mathrm{eq}} = -\frac{H_2 \alpha C P_p \Gamma}{\kappa_{\mathrm{eq}}} e^{\mathrm{j2\omega t}} \tag{S25}$$

Eq. (S25) demonstrates the enhancement provided by the suspended design. As compared to the non-suspended one, the suspension decreases the equivalent thermal conductivity $\kappa_{\mathrm{eq}}$, reducing heat dissipation and contributing to stronger heat accumulation, hence the temperature probed by the probe beam $T_{\mathrm{eq}}$ is larger.

(iii) The temperature change modulates the phase of the probe beam through thermo-optic effect and thermal expansion effect. The amplitude of the 2nd harmonic phase modulation on the probe beam may be calculated based on first order approximation:

$$\Delta\varphi_b = \Delta\tilde{\varphi}_b e^{\mathrm{j2\omega t}} = \frac{2\pi}{\lambda_{\mathrm{b}}} \left(\bar{L}\Delta\tilde{n}_{\mathrm{b}} + \bar{n}_{\mathrm{b}}\Delta\tilde{L}\ \right) e^{\mathrm{j2\omega t}} \tag{S26}$$

here $\lambda_{\mathrm{b}}$ is the wavelength of the probe beam, $\Delta\tilde{n}_{\mathrm{b}}$ and $\Delta\tilde{L}$ are the probe mode RI change and the waveguide length change due to the temperature change. The "–" denotes the time averaged value of the variables. The temperature change modulates the probe mode index change $\Delta\tilde{n}_{\mathrm{b}}$ through thermo-optic effect and thermo-photoelasticity effects by altering the permittivity of the material, while also modulates the waveguide length $\Delta\tilde{L}$ through thermal expansion effect by thermal strain. Probe mode index change $\Delta\tilde{n}_{\mathrm{b}}$ is related to the permittivity change $\Delta\tilde{\varepsilon}_r$ through:

$$\Delta\tilde{n}_{\mathrm{b}} = \sum_i \frac{n_{\mathrm{gb}}}{2n_i} \iint_i \psi_b^* \Delta\tilde{\varepsilon}_{ri} \psi_b\, dxdy \tag{S27}$$

here $n_{\text{gb}}$ is the group index of the probe mode, and the indices $i$ denote the target material (gas/ChG/$SiO_2$/Si). The permittivity change due to the thermal-optic effect is expressed as:

$$\Delta\tilde{\varepsilon}_{ri}^{TOE} = 2n_i \frac{dn_{\text{i}}}{dT}\tilde{T} \tag{S28}$$

$$\Delta\tilde{\varepsilon}_{ri}^{PE} = -n^4 \vec{p}_i : \vec{\tilde{S}} = -n^4 \begin{bmatrix} p_{i11} & p_{i12} & p_{i12} & 0 & 0 & 0 \\ p_{i12} & p_{i11} & p_{i12} & 0 & 0 & 0 \\ p_{i12} & p_{i12} & p_{i11} & 0 & 0 & 0 \\ 0 & 0 & 0 & p_{i44} & 0 & 0 \\ 0 & 0 & 0 & 0 & p_{i44} & 0 \\ 0 & 0 & 0 & 0 & 0 & p_{i44} \end{bmatrix} \begin{bmatrix} \tilde{S}_{xx} \\ \tilde{S}_{yy} \\ \tilde{S}_{zz} \\ \tilde{S}_{zy} \\ \tilde{S}_{xz} \\ \tilde{S}_{yx} \end{bmatrix} \tag{S29}$$

where we have employed $\Delta\tilde{\varepsilon}_r = 2n\Delta\tilde{n}$ for deducting Eq. (S28), $dn_i/dT$ is the thermo-optic coefficient of the target material, $\vec{p}_i$ is the photoelasticity tensor of the target material, $\tilde{S}$ is the amplitude of the strain tensor oscillating at $2\omega_m$. As for the thermal expansion effect, the waveguide length change $\Delta\tilde{L}$ is related to the temperature change $\tilde{T}$ through:

$$\Delta\tilde{L} = \int_{L_0} \frac{\iint_{\text{A}} \mathcal{E}\frac{dL}{dT}\tilde{T}\, dxdy}{\iint_{\text{A}} \mathcal{E}\, dxdy} dz \tag{S30}$$

here $\mathcal{E}$ is the young's modulus of the material, $L_0$ is the length of the waveguide under ambient temperature, and $dL/dT$ is the thermal expansion coefficient of the materials. The subscripts A of the surface integrals denote the integration in the full waveguide cross-section region (gas/ChG/$SiO_2$/Si).

The overall phase modulation is a summation of contributions from TOE, photoelasticity effect and thermal expansion effect as $\Delta\tilde{\varphi}_b = \Delta\tilde{\varphi}_b^{TOE} + \Delta\tilde{\varphi}_b^{PE} + \Delta\tilde{\varphi}_b^{TEE}$, which are accordingly:

$$\Delta\tilde{\varphi}_{\text{b}}^{\text{TOE}} = \frac{2\pi}{\lambda_{\text{b}}} L_0 \Delta\tilde{n}_{\text{b}}^{TOE} \tag{S31}$$

$$\Delta\tilde{\varphi}_{\text{b}}^{\text{PE}} = \frac{2\pi}{\lambda_{\text{b}}} L_0 \Delta\tilde{n}_{\text{b}}^{PE} \tag{S32}$$

$$\Delta\tilde{\varphi}_{\text{b}}^{\text{TEE}} = \frac{2\pi}{\lambda_{\text{b}}} n_{\text{b}} \Delta\tilde{L} \tag{S33}$$

The value of these terms can be numerically solved by combing Eqs. (S31-33) with Eqs. (S27-30). Typically, photoelasticity is extremely weaker than the others since the photoelasticity coefficient is two or three orders of magnitude smaller than the thermo-optic and thermal expansion coefficient. Besides, since the suspended waveguide structure is a multilayer structure, the axial thermal expansion of the waveguide core (ChG) is constrained by mechanical compatibility across other bonded layers ($SiO_2$/Si). Consequently, the thermal expansion efficiency is much lower than that of an isolated, freely expanding ChG waveguide and depends on layer stiffnesses and boundary conditions. Fig. S1 shows the numerically calculated $\Delta\tilde{\varphi}_{\text{b}}^{\text{TOE}}$, $\Delta\tilde{\varphi}_{\text{b}}^{\text{PE}}$ and $\Delta\tilde{\varphi}_{\text{b}}^{\text{TEE}}$, where the relative effort of neglecting the photoelasticity and thermal expansion effect is <0.1%. Hence, the PT phase modulation can be approximated only considering the thermo-optic effect in Eq. (S27-28). Combining Eq. (S27-28) and Eq. (S31), we can solve the probe mode index change $\Delta\tilde{n}_{\text{b}}$ as:

$$\Delta\tilde{n}_{\text{b}} = \frac{n_{\text{gb}}}{n_{\text{ChG}}}\frac{dn_{\text{gas}}}{dT}\iint_{\text{ChG}} \psi_{\text{b}}^* \tilde{T} \psi_{\text{b}}\, dxdy + \frac{n_{\text{gb}}}{n_{\text{gas}}}\frac{dn_{\text{gas}}}{dT}\iint_{\text{gas}} \psi_{\text{b}}^* \tilde{T} \psi_{\text{b}}\, dxdy \tag{S34}$$

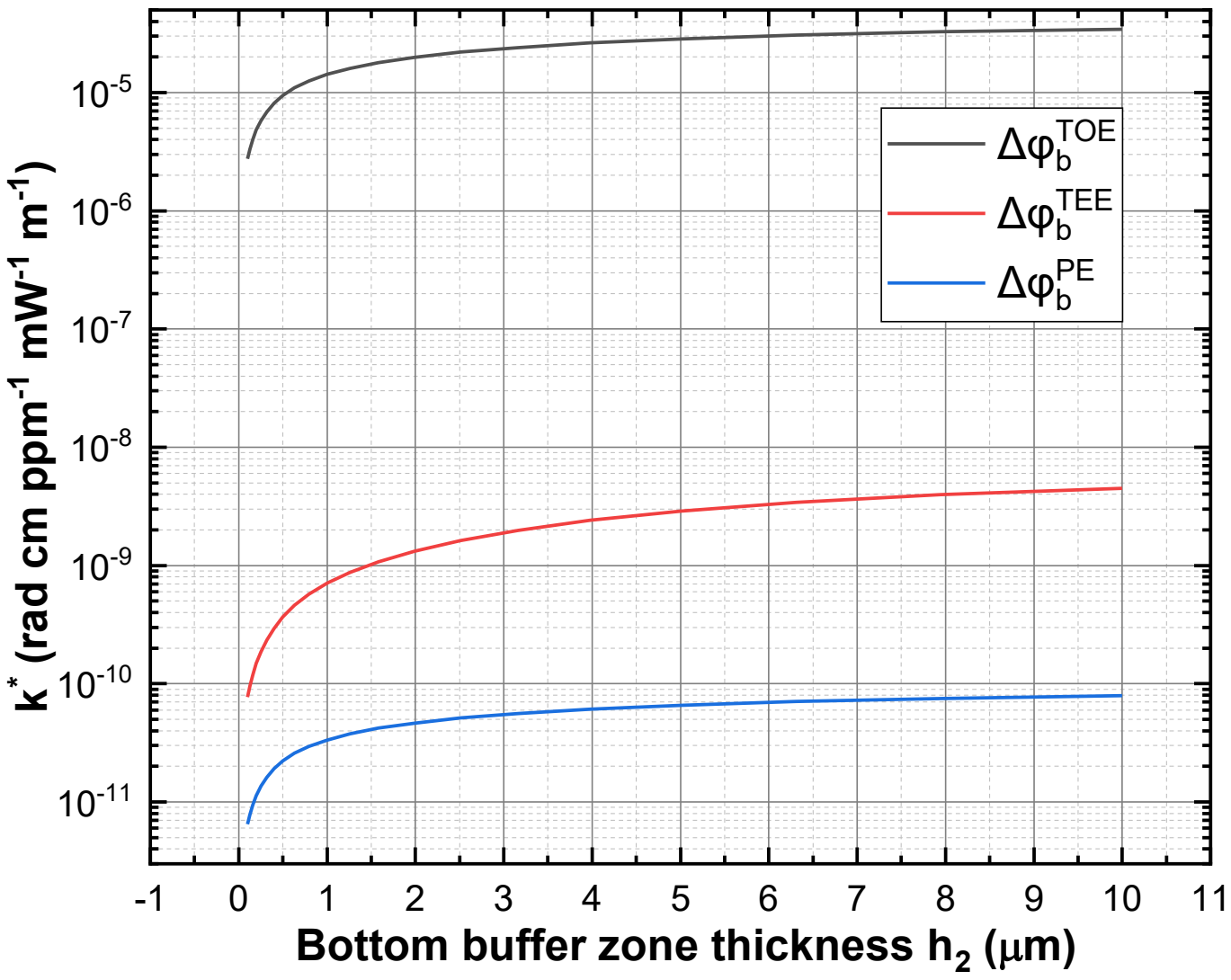

**Fig. S1 Phase modulation efficiency $k^*$ due to thermo-optic effect (TOE), photoelasticity (PE) and thermal expansion effect (TEE).** The phase modulation efficiency is calculated under 100 mW pump power and 100 ppm acetylene for different buffer zone thickness $h_2$.

Following Eq. (S34), we can define an equivalent thermo-optic coefficient $e_{\mathrm{eq}}^{\mathrm{TO}}$ for the waveguide structure:

$$e_{\mathrm{eq}}^{\mathrm{TO}} = \frac{\left(\frac{1}{n_{\mathrm{ChG}}}\frac{dn_{\mathrm{ChG}}}{dT}\iint_{\mathrm{ChG}}\psi_{\mathrm{b}}^{*}\tilde{T}\psi_{\mathrm{b}}\,dxdy + \frac{1}{n_{\mathrm{gas}}}\frac{dn_{\mathrm{gas}}}{dT}\iint_{\mathrm{gas}}\psi_{\mathrm{b}}^{*}\tilde{T}\psi_{\mathrm{b}}\,\mathrm{d}xdy\right)}{\iint_{\mathrm{total}}\psi_{\mathrm{b}}^{*}\tilde{T}\psi_{\mathrm{b}}\,\mathrm{d}xdy}$$

$$= \frac{1}{T_{\mathrm{eq}}}\frac{e_{\mathrm{ChG}}^{\mathrm{TO}}}{n_{\mathrm{ChG}}}\iint_{\mathrm{ChG}}\psi_{\mathrm{b}}^{*}\psi_{\mathrm{T}}\psi_{\mathrm{b}}\,\mathrm{d}xdy - \frac{1}{T_{\mathrm{eq}}}\frac{e_{\mathrm{gas}}^{\mathrm{TO}}}{n_{\mathrm{gas}}}\iint_{\mathrm{gas}}\psi_{\mathrm{b}}^{*}\psi_{\mathrm{T}}\psi_{\mathrm{b}}\,\mathrm{d}xdy \tag{S35}$$

here $e_{\mathrm{ChG,gas}}^{\mathrm{TO}} = \mathrm{d}n_{\mathrm{ChG,gas}}/\mathrm{d}T$ is the thermo-optic (TO) coefficient of waveguide material and gas material. Based on Eq. (S31), (S34) and (S35), we can express the probe phase modulation as a function of $T_{eq}$ as:

$$\Delta\varphi_{\mathrm{b}} \approx \frac{2\pi n_{\mathrm{gb}}L}{\lambda_{\mathrm{b}}}T_{\mathrm{eq}}e_{\mathrm{eq}}^{\mathrm{TO}}e^{\mathrm{j}2\omega\mathrm{t}} = -\frac{2\pi n_{\mathrm{gb}}L}{\lambda_{\mathrm{b}}}\frac{H_{\mathrm{n}}\alpha CP_{\mathrm{p}}\Gamma}{\kappa_{\mathrm{eq}}}e_{\mathrm{eq}}^{\mathrm{TO}}e^{\mathrm{j}2\omega\mathrm{t}} \tag{S36}$$

Our approximation leads to a linear dependence of the probed temperature $T_{\mathrm{eq}}$ on the power of the heat source $P_Q$, which is consistent with the experimental result shown in Fig. 6(c) of the Main Manuscript.

As depicted by Eq. (S36), the probe phase modulation $\Delta\varphi_{\mathrm{b}}$ and hence the PTS signal depends on the dimensions of the core-layer structure via heat conduction $\kappa_{\mathrm{eq}}$ and hence probed temperature $T_{\mathrm{eq}}$, as well as the core-layer material that decides the thermo-optic coefficient $e_{\mathrm{eq}}^{\mathrm{TO}}$. To compare the PTS performance among different suspended geometric parameters, we define a normalized PT phase modulation efficiency as:

$$k^{*}(w,h,k) \coloneqq \frac{\Delta\varphi_{\mathrm{b}}}{\alpha CP_{\mathrm{p}}L} \propto \Gamma\frac{e_{\mathrm{eq}}^{\mathrm{TO}}}{\kappa_{\mathrm{eq}}} \tag{S37}$$

We numerically calculated the $k^*$ for different parameters using finite element method, and corresponding results are drawn in Fig. 2 of the Main Manuscript.

Note 2. On-chip Fabry-Pérot (F-P) cavity interferometer

The design principle of the F-P cavity is to achieve high fringe contrast, thereby enabling high phase-to-intensity conversion efficiency. This requires that the first reflection at the input facet and the second reflection from the output facet returning to the input facet have comparable power. Relevant parameters include the end-facet coupling loss, transmission loss, reflections between the interface of air/waveguides, and the waveguide length. To achieve this, we employed end-face polishing to minimize scattering at the waveguide facets. Tapered fibers with small mode-field diameter (~500 nm) were carefully aligned to the waveguide facets to reduce the coupling loss. In addition, a two-step patterning process was applied to fabricate the low-loss suspended ChGW. Fig. S2 illustrates the schematic and configuration of an on-chip F-P cavity interferometer. The reflection of the probe beam at two interfaces are governed by the Fresnel reflection law, which are formed between the air (*n*=1) and ChGW (*n*=2.7) with $R_1 = R_2 \approx 21\%$. The optimization of the waveguide length is demonstrated in Supplementary Note 3.

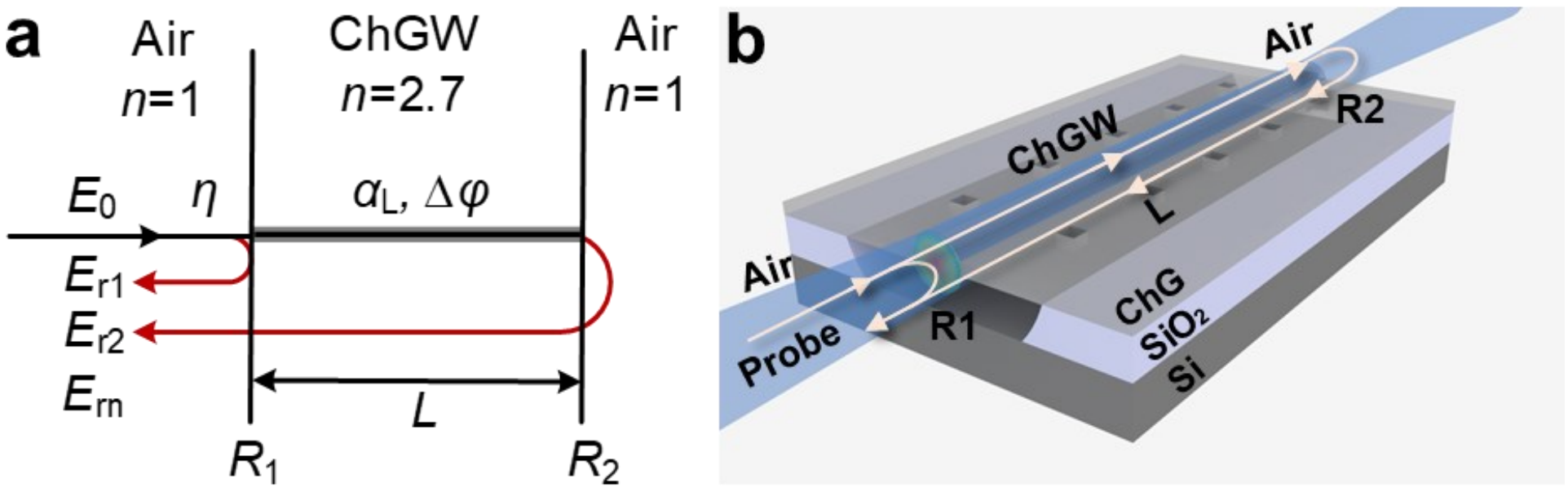

**Fig. S2 On-chip F-P cavity model. (a)** This model uses inherent Fresnel reflections between the interface of air/waveguides. $\eta$ is the end-facet coupling loss, $\alpha_{\mathrm{L}}$ the transmission loss along the ChGW length $L$, $\Delta\varphi$ the probe phase modulation. **(b)** Schematic of the F-P cavity interferometer with suspended ChGW.

Assuming the input probe electric field is $E_0$, the reflected field at the first interface can be represented as $E_{\mathrm{r1}} = E_0\sqrt{R_1\eta}$. The transmitted probe beam undergoes coupling loss ($\eta$), transmission loss ($\alpha$), and PT-induced phase modulation $\Delta\varphi$. After traversing the waveguide length $L_{\mathrm{WG}}$, the light reflects a second time as $E_{\mathrm{r2}} = E_0(1-R_1)\sqrt{R_2\eta}e^{2\alpha\mathrm{L}}e^{\mathrm{j}\varphi}$. While more reflections ($E_{\mathrm{rn}}$) could occur, losses beyond the third reflection become large enough, thus such higher-order reflections are generally negligible. Consequently, the reflected signal resulting from the F-P cavity can be described as

$$I = (E_{\mathrm{r1}} + E_{\mathrm{r2}})(E_{\mathrm{r1}} + E_{\mathrm{r2}})^* = E_0^2 R_1\eta + E_0^2 m^2 + 2E_0^2\sqrt{R_1\eta}me^{\mathrm{j}\varphi} \tag{S38}$$

where $m$ is defined as $m = (1-R_1)\sqrt{R_2\eta}e^{2\alpha\mathrm{L}}$. Here the phase difference $\varphi$ between the two reflected beams is expressed as

$$\varphi = \frac{4\pi n_{\mathrm{gb}} L_{\mathrm{WG}}}{\lambda_{\mathrm{b}}} + \varphi_0 \tag{S39}$$

here $\varphi_0$ is the initial phase of the suspended ChGW. The fringe contrast $\upsilon$ can be expressed as

$$\upsilon = \frac{I_{\max} - I_{\min}}{I_{\max} + I_{\min}} = \frac{2m\sqrt{R_1\eta}}{R_1\eta + m^2} \tag{S40}$$

Assuming a sinusoidal phase modulation $\varphi = \Delta\varphi \sin(\omega t) + \varphi_0$ with pump modulated angular frequency $\omega$, the normalized interferometric signal can be expressed in the Bessel function as:

$$I_{\text{norm}} = R_1\eta + m^2 + 2m\sqrt{R_1\eta}\,\{\cos(\varphi_0)[J_0(\Delta\varphi) + 2\sum_{n=1}^{\infty} J_{2n}(\Delta\varphi)\cos(2n\omega t)]$$

$$+\sin(\varphi_0)[2\sum_{n=0}^{\infty} J_{2n+1}(\Delta\varphi)\sin((2n+1)\,\omega t)]\} \quad \text{(S41)}$$

where $J_n(\Delta\varphi)$ is the Bessel function of order *n* with argument of $\Delta\varphi$. In this manner, the phase modulation could be converted into intensity output of the interferometer after the photodetector. The amplitude of $\Delta\varphi$ is always of main interest. If we consider Eq. (S41) with weak absorption approximation: $\Delta\varphi \ll 1$, $J_0 = 1$, $J_1(\Delta\varphi) = \Delta\varphi/2$, the reflected intensity could be simplified as:

$$I_{\text{norm}} = 1 + \frac{2m\sqrt{R_1\eta}\sin(\varphi_0)}{R_1\eta + m^2}\Delta\varphi\sin(\omega t) \quad \text{(S42)}$$

In this setup, the F-P cavity is working at a quadrature point, that is $\varphi_0 = \pi/2$. The output intensity could be expressed as

$$I_{\text{norm}}(t) = R_1\eta + m^2 + 2m\sqrt{R_1\eta}\,\Delta\varphi\sin(\omega t) = 1 + v\Delta\varphi\sin(\omega t) \quad \text{(S43)}$$

The PTS signal amplitude ($S_{\text{PTS}}$) is defined as the 2*f*-demodulated voltage amplitude, which is proportional to the average probe power ($P_{\text{avg}}$) at the PD, $v$, and $\Delta\varphi$, expressed as

$$S_{\text{PTS}} \propto P_{\text{avg}}\, v\, \Delta\varphi \quad \text{(S44)}$$

## Note 3. Optimization of the suspended waveguide length

The accumulated PT-induced phase modulation $\Delta\varphi$ generally increases with the waveguide length *L*. However, with increasing *L*, the fringe contrast $v$ of such an F-P cavity also diminishes as the intensity of the light reflected from the second end of the ChGW decreases due to increasing transmission loss $\alpha_L$ along the waveguide, resulting in the degradation of $S_{\text{PTS}}$. Therefore, there is a trade-off between $\Delta\varphi$ and $v$ with an optimal chip length *L*. As shown in Fig. S3(a), the transmission loss of the suspended ChGW is measured to be 2.6 dB/cm. The power *P* along the waveguide length *L* is expressed as

$$\log_{10}\frac{P(L)}{P_0} = \frac{\alpha_L \cdot L|_{\text{dB}}}{10} \quad \text{(S45)}$$

here $P_0$ is the initial power coupled into the suspended ChGW. The normalized fringe contrast $v_{\text{norm}}$ is in the form of:

$$v_{\text{norm}} = \frac{2m\sqrt{R_1\eta}}{R_1\eta + m^2} = \frac{2\sqrt{R_1\eta}(1-R_1)\sqrt{R_2\eta}\,e^{2\alpha_L L}}{R_1\eta + [(1-R_1)\sqrt{R_2\eta}\,e^{2\alpha_L L}]^2} \quad \text{(S46)}$$

The PT phase modulation $\Delta\varphi$ could be calculated as a function of *l*:

$$\Delta\varphi = \frac{2\pi}{\lambda_b}\int_0^L k^*\,P(L)dl \quad \text{(S47)}$$

According to Eq. (S45), the $S_{\text{PTS}}$ is proportional to the product of $\Delta\varphi$ and $v$. The results of $\Delta\varphi$, $v$ and their combined effect on $S_{\text{PTS}}$ in terms of *L* is presented in Fig. S3(b). Calculations indicate that the optimal waveguide length for achieving the maximum PTS signal is approximately 0.8 cm, maintaining over 80% of the maximum value within the 0.5−1.4 cm range.

Although the optimal waveguide length is 0.8 cm, we need to encapsulate it within a gas chamber and provide space for the gas inlet and outlet. As a result, we select a slightly longer length of 1.2 cm. At this length, the PTS performance still maintain about 90% of that at the optimal length. Additionally, a portion of the waveguide extends beyond the gas chamber to facilitate easier fiber coupling. The effective waveguide length interacting with gas inside the chamber is ~0.8−1 cm.

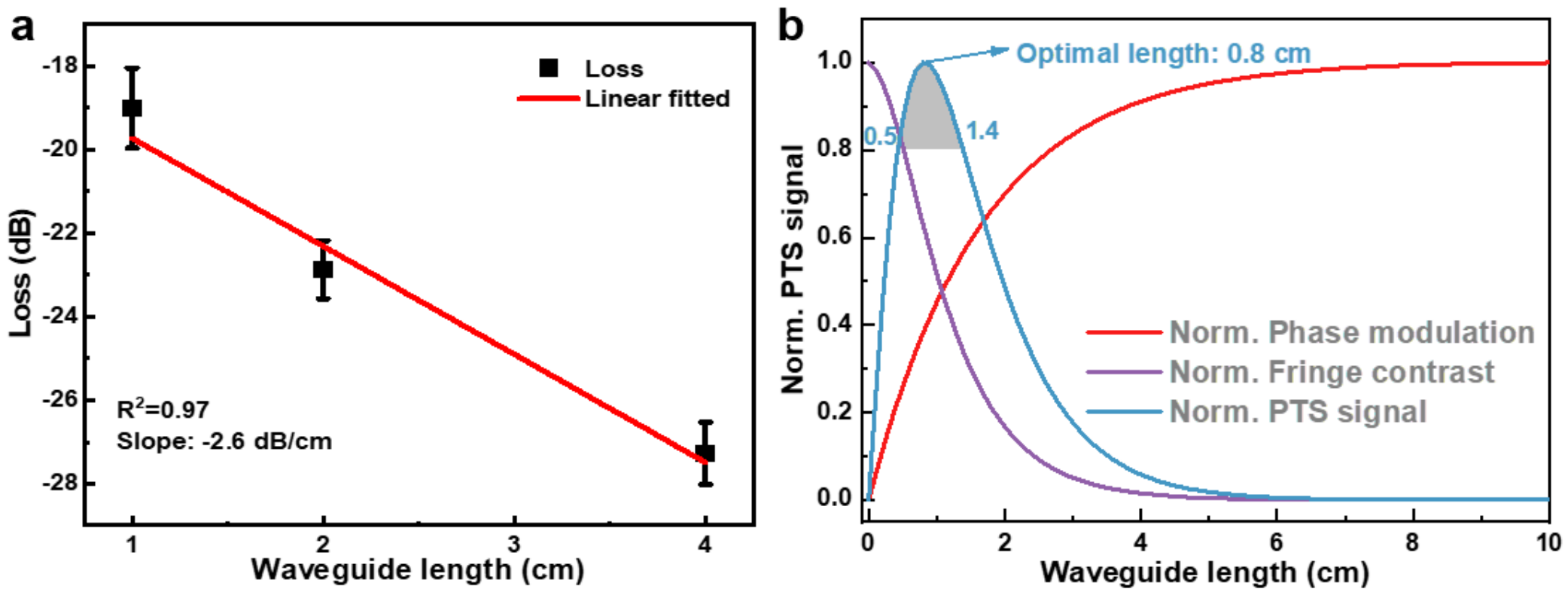

**Fig. S3 Suspended ChGW loss assessment and length optimization. (a)** Measured transmission loss of the suspended ChGW using different waveguide lengths. **(b)** Optimization of the suspended ChGW length to achieve the maximum PTS signal amplitude.

## Note 4. Comparison between suspended and non-suspended ChGW

For the suspended ChGW, COMSOL Multiphysics is utilized for numerical simulation with parameters designed as $w$=1000 nm, $h$=300 nm, $k$=0.5, as shown in Fig. S4(a). Here suspended geometric parameters includes the rib width ($w$), height ($h$) and rib ratio ($k$). The heat source distributions are shown in Fig. S4(b) with different photothermal dynamics under varying demodulation frequencies. As demodulation frequency $f$ increases, a concentration of heat towards the core layer with a corresponding decrease in the heat field is observed. This reduction is mainly attributed to the decreased heat generation within a modulation period. Fig. S4(c) depicts the probed temperature change in heat production in time domain, showing a decrease in steady-state temperature change from 13.7 to 1.26 mK as frequency increases from 2 to 200 kHz.

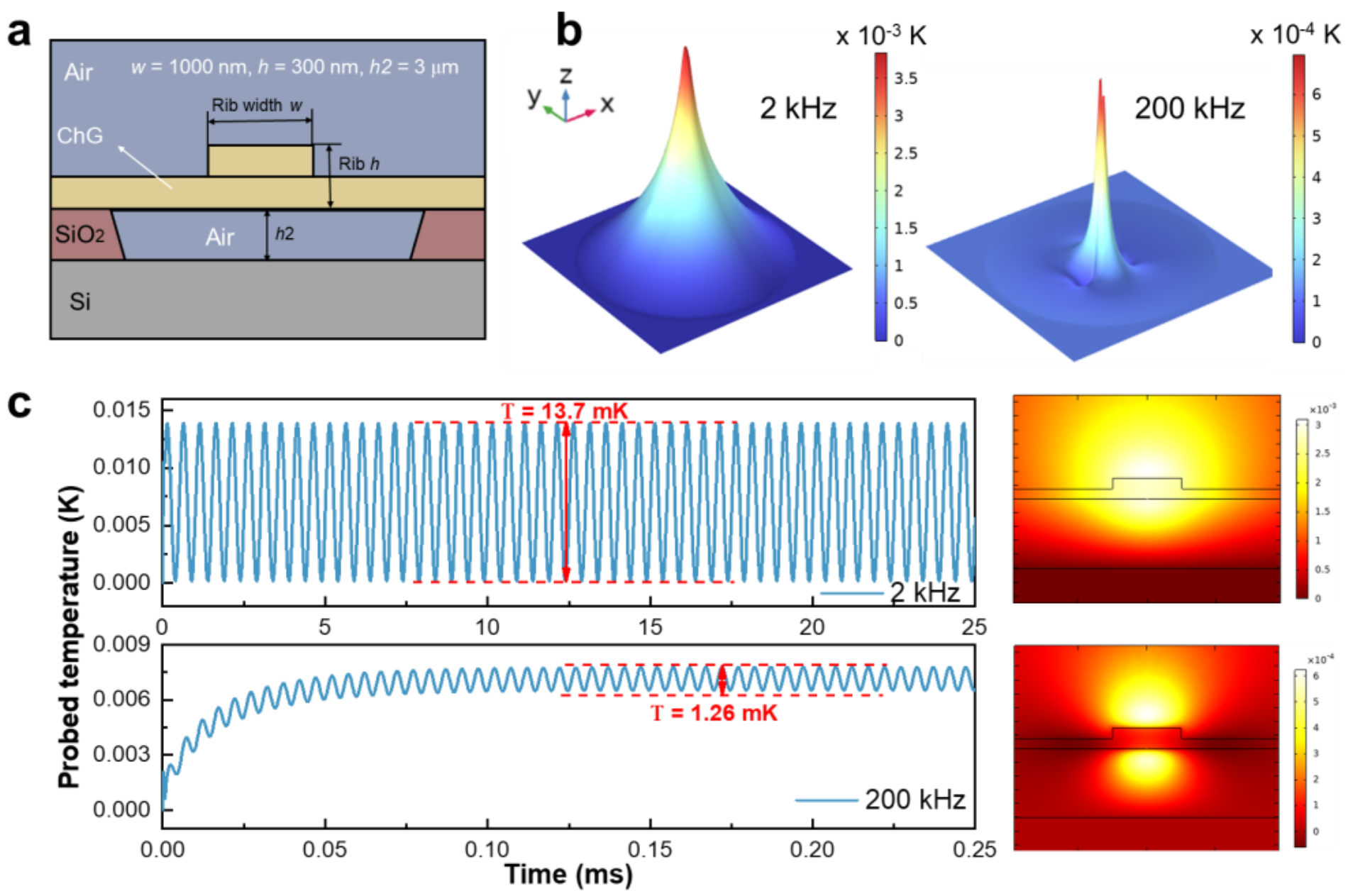

**Fig. S4 Suspended ChGW PT dynamics. (a)** Suspended ChGW model and **(b)** heat source profile with different demodulation frequency (2 and 200 kHz), respectively. **(c)** Dynamics of the heat conduction in time domain with corresponding steady-state thermal profiles.

Following a similar analytical approach in Fig. S4, we investigate the PT dynamics of non-suspended ChGW using parameters specified as $w$=1000 nm, $h$=300 nm, $k$=1 with $SiO_2$ as the bottom cladding, not the air. As shown in Fig. S5(a). The heat source profiles with varying $f$ are shown in Fig. S5(b), while Fig. S5(c) represent the time-domain dynamics, and the steady-state probed temperature decreases from 0.33 to 0.21 mK as $f$ increases from 2 to 200 kHz.

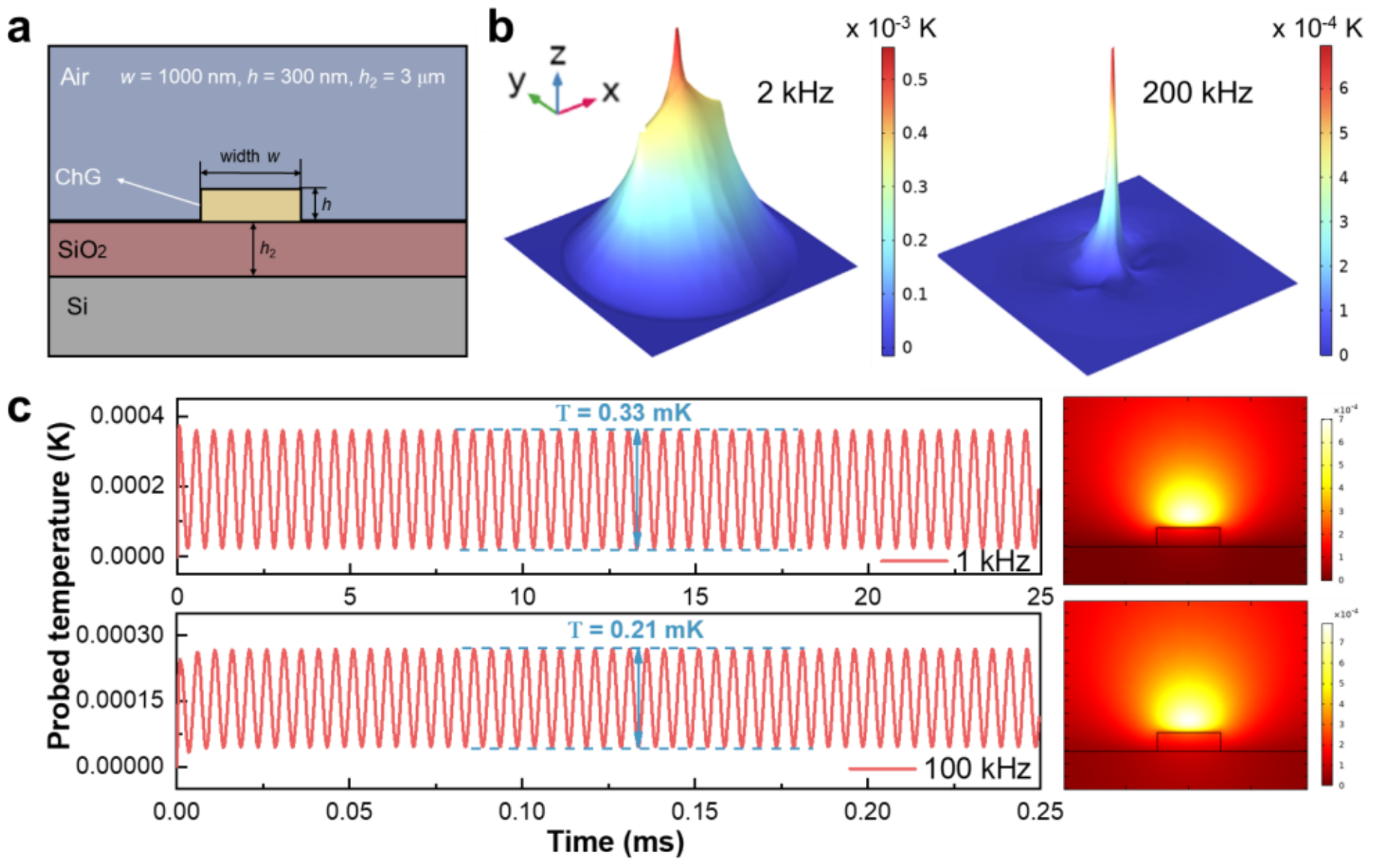

**Fig. S5 Non-suspended ChGW PT dynamics. (a)** Non-suspended ChGW model and **(b)** heat source profile with different demodulation frequency (2 and 200 kHz), respectively. **(c)** Dynamics of the heat conduction in time domain with corresponding steady-state profiles.

Detailed performance comparisons between these two waveguides are provided in Table S1. It is clearly seen that the GCF ($\Gamma$) of the suspended waveguide is nearly 4 times larger than that of the non-suspended one, yet it exhibits 10.6 times slower heat conduction and 1.08 times larger equivalent thermo-optic coefficient. Moreover, the heat source-induced probed temperature $T_{\mathrm{eq}}$ of the suspended waveguide is ~42 times larger than that of the non-suspended. The overall $k^*$ enhances 45 times at 2 kHz, which further verifies the suspended ChGW a better option for on-chip PTS than traditional non-suspended strip one.

**Table S1** Detailed comparisons of the suspended- and non-suspended waveguide

| Waveguide structures | $\Gamma$ (%) | $\kappa_{\mathrm{eq}}$ (W/(m·K)) | $e_{\mathrm{eq}}^{\mathrm{TO}}$ (1/K) | $T_{\mathrm{eq}}$ (mK) | $k^*$ ($10^{-6}$ rad cm $\mathrm{ppm}^{-1}$ $\mathrm{mW}^{-1}$ $\mathrm{m}^{-1}$) @2/20/200 kHz |
|---|---|---|---|---|---|
| **Suspended** | 95 | 0.075 | $3.32\times10^{-6}$ | 4.71 | 19/9.3/1.7 |
| **Non-suspended** | 24 | 0.793 | $3.07\times10^{-6}$ | 0.112 | 0.42/0.4/0.24 |
| **Enhancement** | **4** | **10.6** | **1.08** | **42** | **45/23/7** |

## Note 5. Suspended ChGW fabrication

The suspended ChGW core layer comprises a bottom slab and a top rib waveguide. The ratio of the thickness of the bottom slab to the rib waveguide should be more than 1:4 to ensure its robustness and stability. The fabrication of suspended waveguide is more complex than traditional channel waveguide, which requires two DUV patterning. Introducing a silicon oxide intermediate layer is more advantageous for removing photoresist from ChGW. In addition, the process involves the use of HF wet etching techniques to remove the underlying silicon oxide layer, demanding high precision to prevent ChGW from crack.

We utilize a 3-μm-thick thermally oxidized silicon wafer, a material that is harder and reacts more slowly with HF than other CVD-grown silicon oxides, aiding in the precise control of the etching rate. As shown in Fig. S6, the fabrication process is divided into two stages: initially, after depositing the required ChG and $SiO_2$ films on the wafer, the first pattern is performed to create micro-holes for HF wet etching. The square holes, measuring 3×3 $\mu m^2$ and spacing 7 μm apart, are patterned by using a Canon DUV stepper lithography system, followed by ICP etching. The structure is cleaned using NMP (1-methyl-pyrodine) solvent and IPA, then remove the silica intermediate layer with HF, and finally ultrasonically clean for 30 s to remove the residue completely. The second stage involves the precise alignment and patterning of the ChG rib waveguide over the first layer. DUV alignment error <40 nm ensures high precision of secondary lithography, the rib waveguides are etched to the designed thickness using ICP and then the photoresist is stripped, exposing the clean ChG rib waveguide. The final step involves immersing the sample in 5% HF, initiating wet etching of the thermally oxidized silicon through the micro-holes fabricated in the first stage. This etching process is meticulously timed and controlled to prevent over-etching and structural damage. After processing, the samples are soaked and rinsed in DI water to completely remove any residual HF.

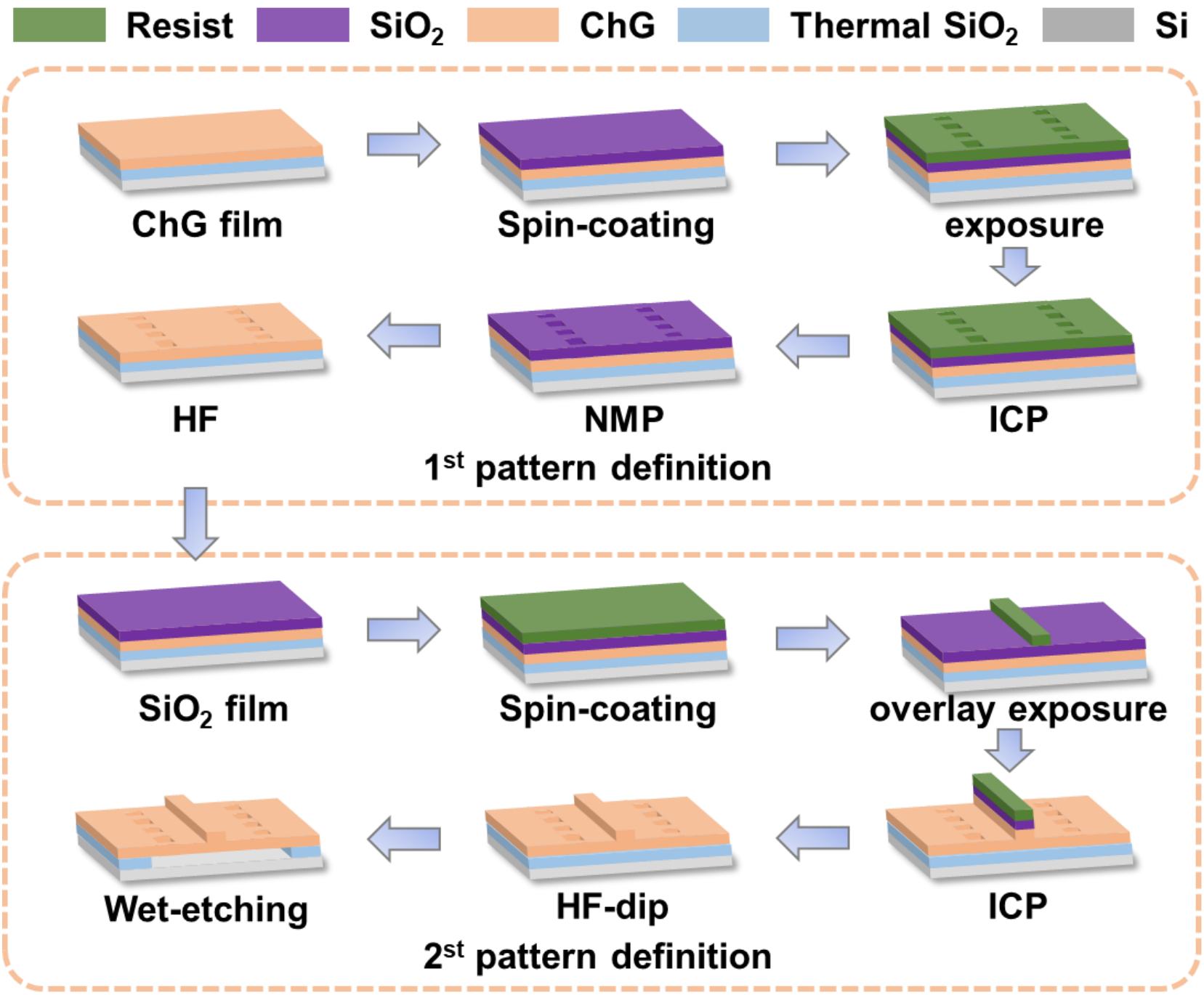

**Fig. S6 Fabrication process of suspended ChGW with two-step patterning process.** ICP: inductive-coupled plasma, NMP: 1-methyl-pyrodine.

## Note 6. Suspended ChGW characterization

As shown in Figs. S7(a)-(c), controlling the wet-etching time allows for precise management of the lateral and vertical dimensions of the etched silicon oxide. As the HF etching time increases, the 3-μm-thick silicon layer beneath the ChG waveguide is more thoroughly removed, eliminating the silicon in the vertical direction while gradually increasing the lateral etching depth. Particularly, at an etching time of 50 min, the silicon oxide under the suspended waveguide is entirely removed. The integrity of the structure can be confirmed by scanning electron microscope (SEM) images. Inset of Fig. S7(c) includes a microscopic image of the suspended structure, visually illustrating the etching effects. Due to differences in transparency between the etched and unetched areas, the sides of the waveguide appear lighter, making the regions near the membrane and micro-holes visibly brighter than other areas, which is consistent with SEM observations. The CMOS-compatible process using a two-step stacked DUV lithography method ensures the accuracy of the pattern, and the design of the rib waveguide structure size endows the suspended waveguide with excellent structural stability.

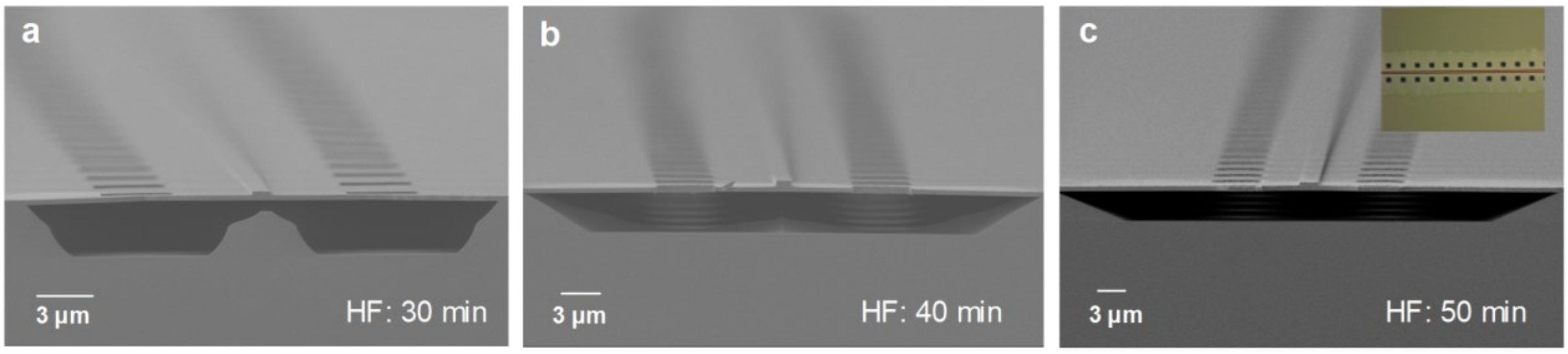

**Fig. S7 Suspended ChGW characterization.** SEM image under different etching time **(a)** 30 min **(b)** 40 min **(c)** 50 min. Inset is a microscopic top view of the suspended ChGW.

Since the quality of the film directly impacts the waveguide loss, it is hence crucial to monitor and characterize the film performance. A Zeiss SEM is used to capture surface and thickness images, while an energy-dispersive spectrometer (EDS) analyzes the film's chemical composition. An atomic force microscope (AFM) characterizes the film's roughness. The deposition rate significantly affects the film's quality. As shown in Fig. S8, under clean chamber conditions, the AFM and SEM images of the ChG films deposited at 25 Å/s are displayed in Fig. S8(a). At this rate, the films exhibit considerable roughness with an average roughness ($R_q$) of 0.9 nm due to the rapid aggregation of gaseous molecules during the growth process, which leads to uneven surfaces. Reducing the current to decrease the deposition rate to 16 Å/s and 13 Å/s results in films with smoother surfaces, as shown in Figs. S8(b) and (c), with $R_q$ values of 0.529 nm and 0.476 nm are obtained, respectively.

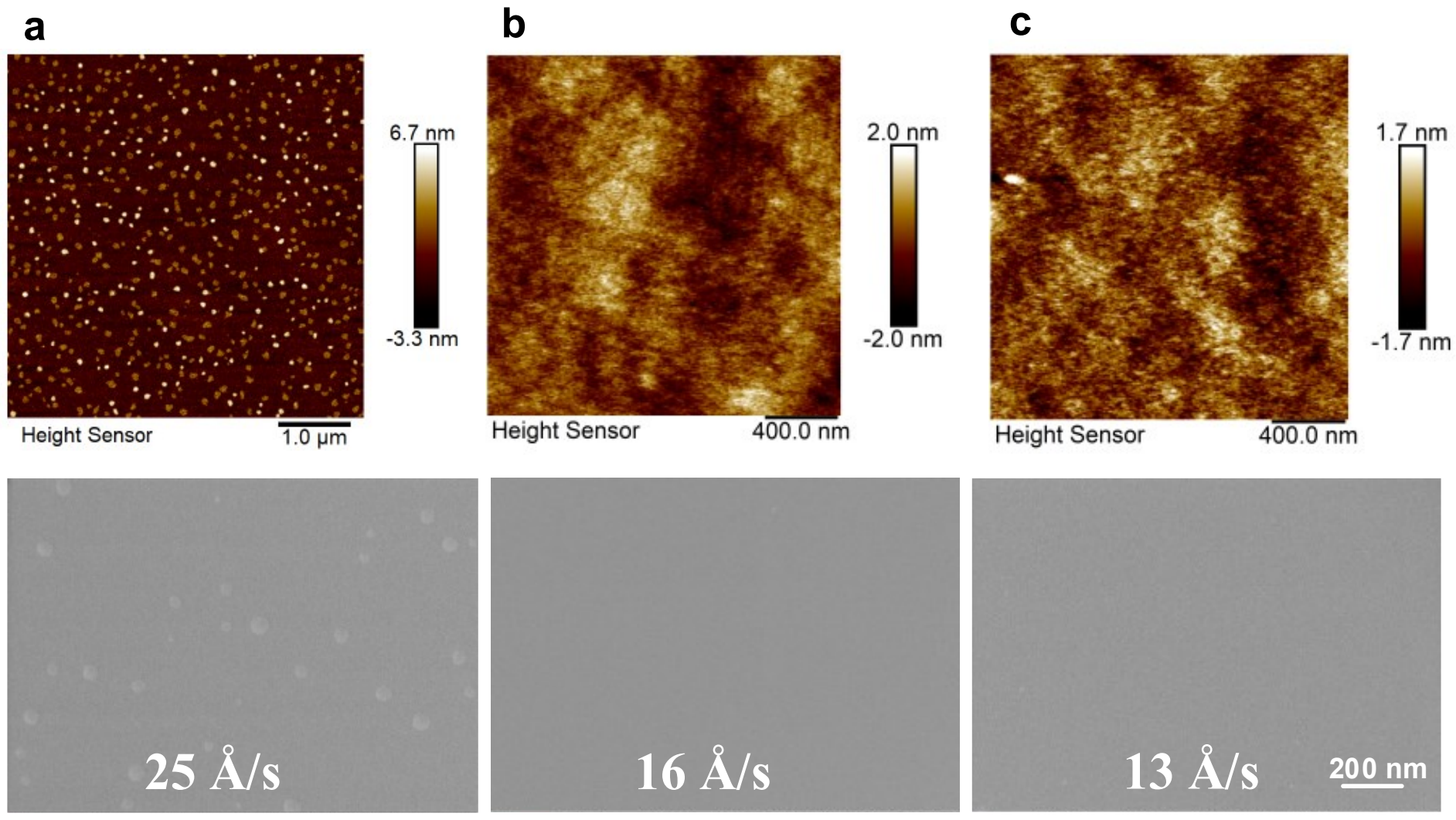

**Fig. S8 ChG film characterization.** AFM and SEM images of films deposited at different rates of **(a)** 25 Å/s, **(b)** 16 Å/s, and **(c)** 13 Å/s.

We then conduct EDS analyses on wafer with deposition rates of 13 Å/s. Fig. S9 shows the EDS mapping for a $Ge_{28}Sb_{12}Se_{60}$ film, revealing a uniform distribution of Ge, Sb, and Se elements. It also presents the energy spectrum analysis of the ChG film grown at 13 Å/s, where peaks corresponding to C, O, and Si are attributed to the substrate elements, and the most prominent peaks are due to Se, indicating its higher proportion in this film. Through composition analysis, the measured element ratios of Ge, Sb, and Se are almost consistent with theoretical values of 28:12:60.

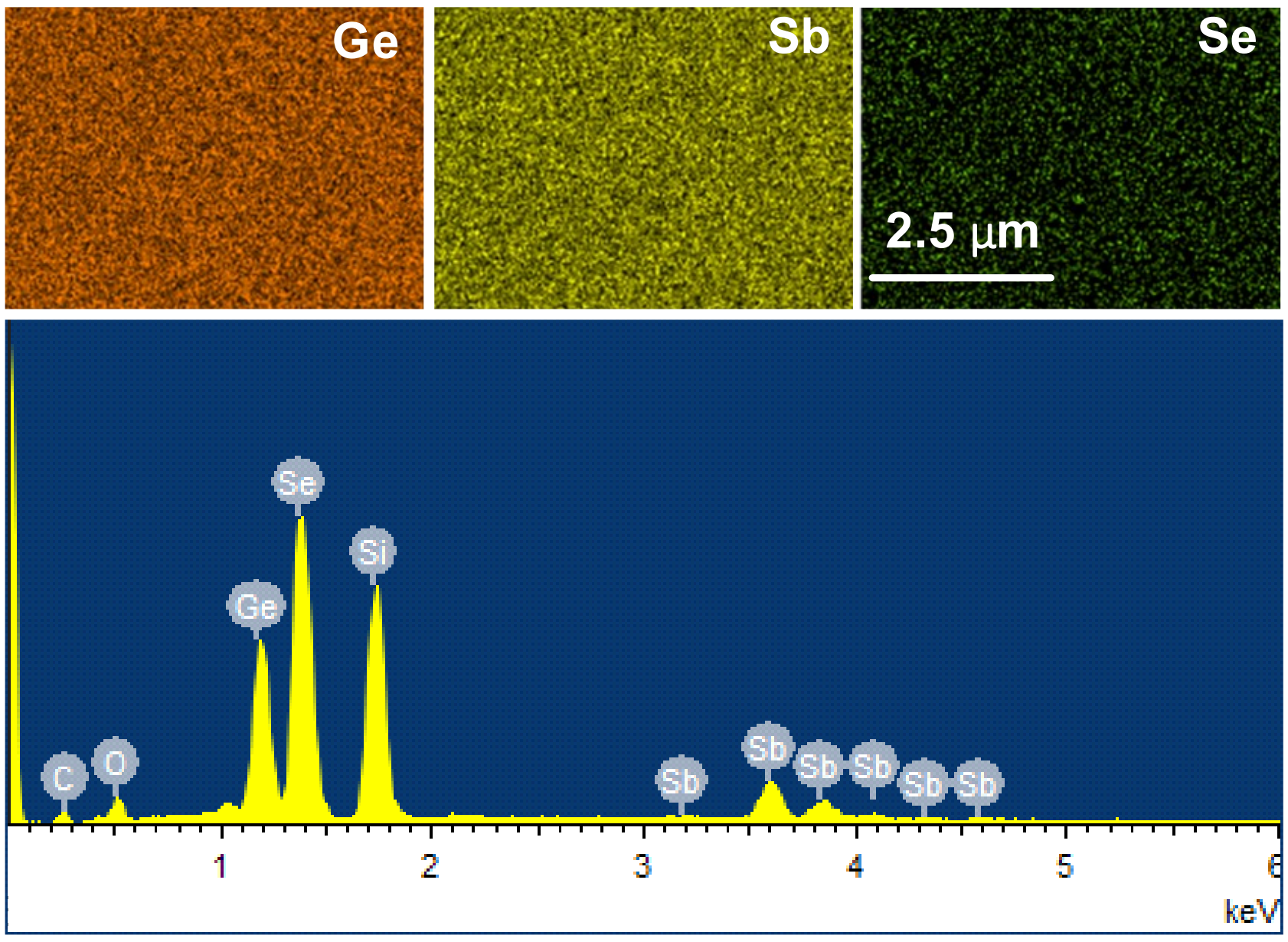

**Fig. S9 $Ge_{28}Sb_{12}Se_{60}$ film characterization.** EDS image of the films deposited at rate of 13 Å/s.

RI is one of the crucial optical parameters of materials. In this work, an ellipsometer, which measures film thickness and optical constants of material films, is used for RI measurement. We employ a German Sentech ellipsometer with a testing wavelength range of 780 nm to 2500 nm. The Sellmeier dispersion model is used for fitting to ensure a fit accuracy better than 99.99%. The resulting RI curve is shown in Fig. S10, where a RI value of ~2.75 is obtained around 1.5 μm.

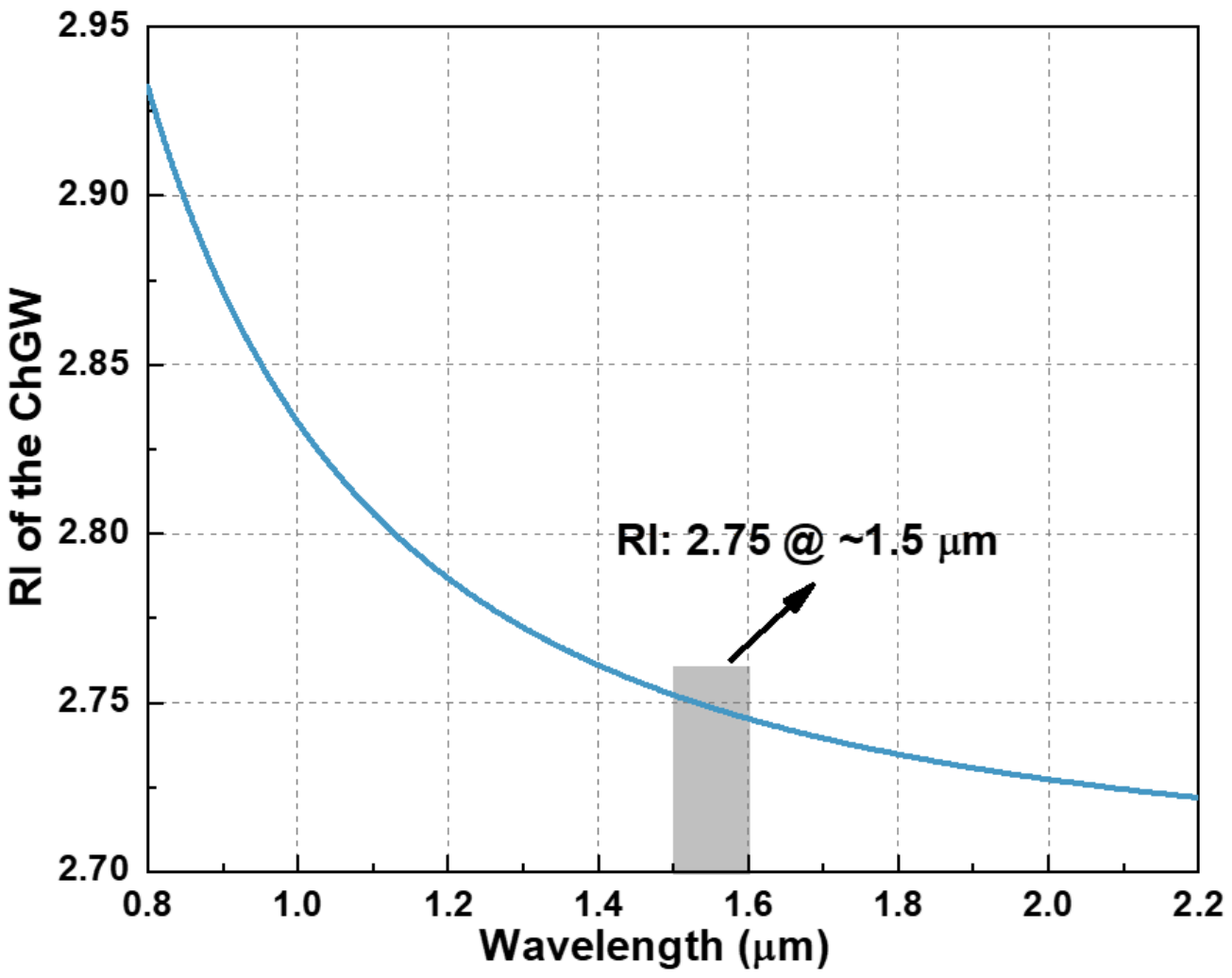

**Fig. S10 Measured RI curve of the ChG film.** The wavelength range is from 0.8 to 2.2 μm.

## Note 7. Experimental investigation of the suspended ChGW

The demonstrated ChGW could measure $C_2H_2$ gas at different absorption lines, as shown in Fig. S11, we coupled a broadband amplified spontaneous emission (ASE) source into the suspended waveguide and collected the transmitted beam using an optical spectrum analyzer (OSA). The $C_2H_2$ absorption spectrum was achieved after filling the pure $C_2H_2$ gas into the waveguide-integrated gas chamber. Each valley corresponded to different $C_2H_2$ absorption lines. Among them, the P(11) absorption line at 1531.58 nm was selected as the pump wavelength for the on-chip PTS, as it exhibited the strongest absorption coefficient in the P-branch of the near-infrared region. In addition, using pump lasers of different wavelengths, we could measure other absorption lines and gas species, as long as the pump wavelength falls within the waveguide's transmission range.

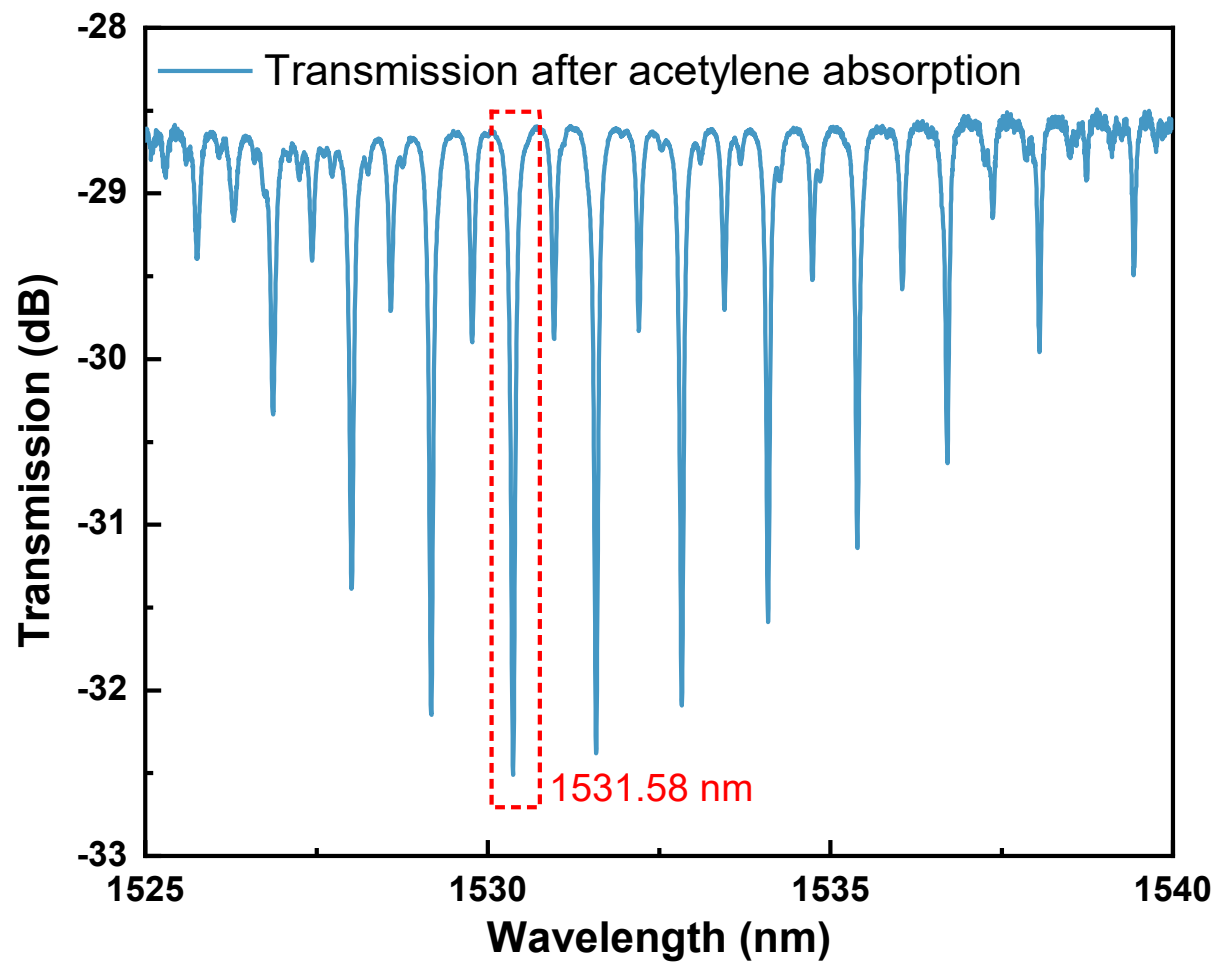

**Fig. S11** The transmission spectrum of ChGW after acetylene absorption.

Fig. S12 shows the reflection spectrum of the F-P interferometer around the pump and probe wavelengths. The excitation of the TM mode primarily relies on the offset alignment between the input fiber and the waveguide. By observing the interference fringes during the alignment process, we could determine whether the excitation is primarily TE or TM. After alignment, as the mode fields of TE and TM differ significantly, adjusting the polarization controllers (PCs) could not affect the pump/probe mode field and fringe spacing, but affect the fringe contrast.

The transmission loss of fundamental mode is small, resulting in larger fringe contrast, while higher-order modes have larger losses, leading to smaller contrast. As seen in Fig. 5(b), the excitation primarily involves the fundamental TM mode, with a very small contribution from higher-order modes. Although ChGW may support higher-order modes, they were not significantly excited in this case. Regarding the effects of higher-order modes on PT efficiency, higher-order pump modes are beneficial for PTS as they generate a larger evanescent field and higher heat source power, but may suffer from larger transmission losses. In contrast, higher-order probe modes should be avoided, as they experience weaker PT effects due to the thermo-optic effect being dominated by ChG rather than air. Higher-order modes have more optical power distributed in air, which results in lower PT efficiency. Additionally, these modes incur higher losses, leading to reduced fringe contrast.

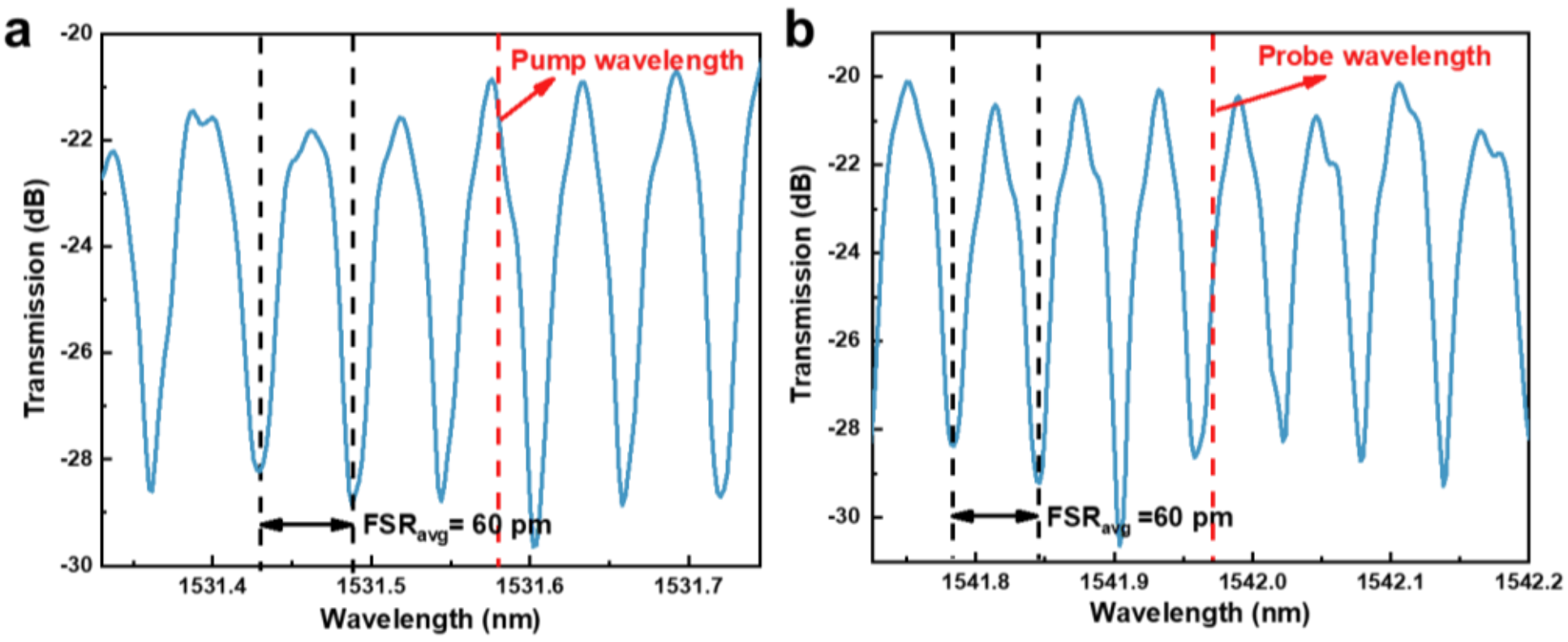

**Fig. S12** Reflection spectrum of the F-P cavity around the **(a)** pump and **(b)** probe wavelengths.

## Note 8. Numerical investigation of tracing different gases

The principle of the SWE-PTS is to measure the heat generated by molecular absorption and the resulting phase modulation, which can be employed for tracing multiple gas species as long as the sensing platform supports multiple absorption wavelengths. We first calculated the leakage loss of the TM mode in the suspended ChGW over the wavelength range from ~1000 to ~3000 nm using the finite element method in COMSOL. We found in Fig. S13 that the fabricated ChGW supports low-loss transmission over the wavelength range from 1000 to 2500 nm. Therefore, we carried out photothermal simulations for some representative gas species with absorption in this wavelength range. We have simulated the photothermal phase modulation for four gas species, i.e., $C_2H_2$, $CH_4$, CO and $CO_2$ using pump wavelengths corresponding to their relatively strong absorption lines at 1.531 μm, 1.651 μm, 1.568 μm, and 2.004 μm, respectively. The thermo-optic and thermodynamic parameters used in the simulation are summarized in Table S2.

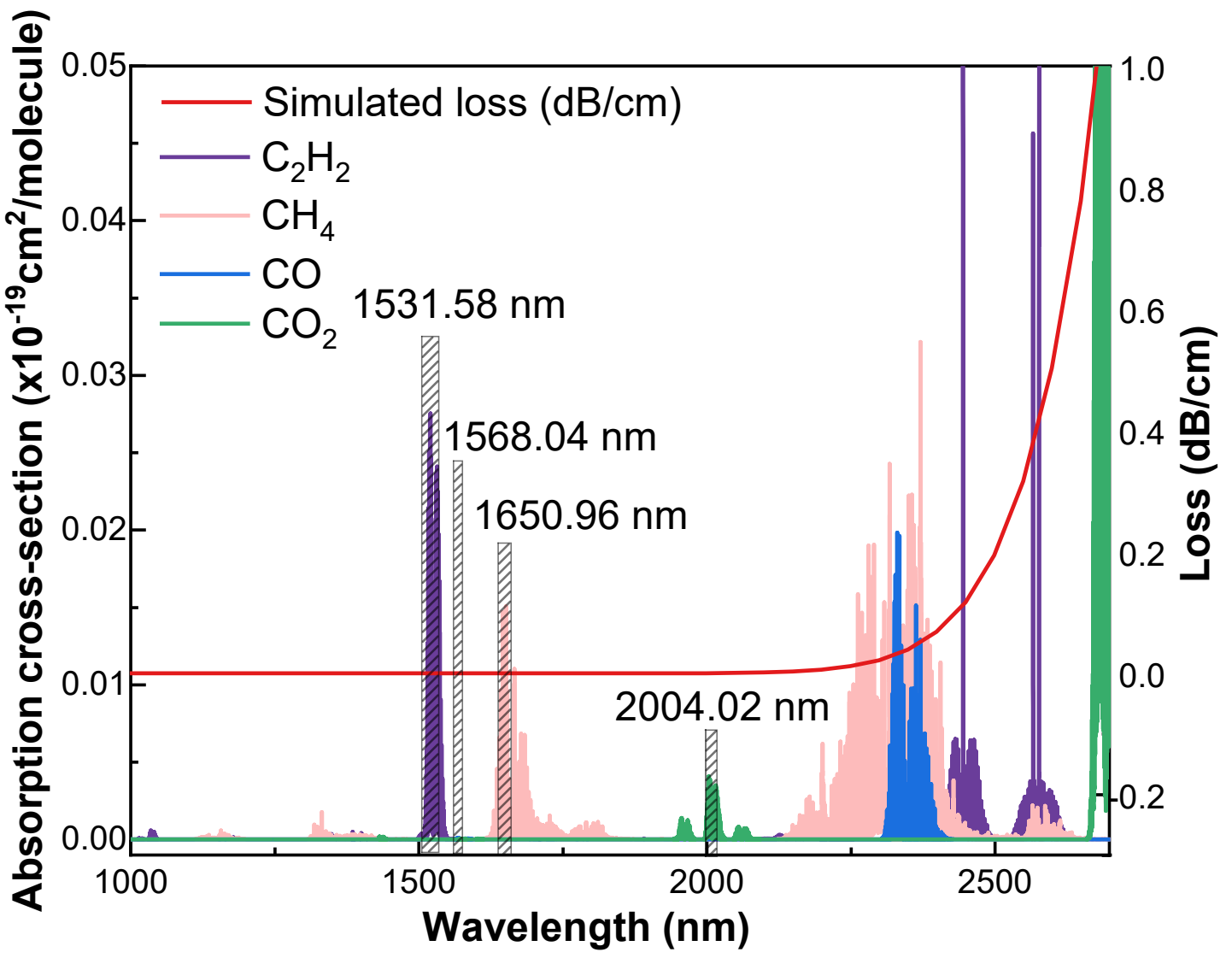

**Fig. S13** Simulated leakage loss of the suspended ChGW and the absorption cross-sections of the four gas species ($C_2H_2$, $CH_4$, CO and $CO_2$).

**Table S2** Parameters used in this simulation

| Gas type | Wavelength (nm) | Rovib mode | $\alpha$ (cm$^{-1}$) | $\tau_{re}$ (μs·atm) | Evanescent proportion ($\gamma_p$) | Pump group index ($n_{gp}$) | Gas confinement factor ($\Gamma$) |
|---|---|---|---|---|---|---|---|
| $C_2H_2$ | 1531.58 | $\nu_1+\nu_3$ | 1.050 | 0.074 | 30.4% | 3.15 | 95% |
| $CH_4$ | 1650.96 | $2\nu_3$ | 0.453 | 3.8 | 40.7% | 2.76 | 112% |
| CO | 1568.04 | $\nu_3$ | 0.003 | 25 | 33.8% | 3.04 | 102% |
| $CO_2$ | 2004.02 | $2\nu_1+\nu_3$ | 0.142 | 11 | 45.6% | 2.19 | 99% |

Rovib mode: rovibrational mode of the molecule, $\alpha$: absorption coefficient, $\tau_{re}$: relaxation time constant of the target rovibrational mode, $\Gamma$ is the gas confinement factor that scales the heat source power due to pump absorption, which is equal to $n_{gp}\gamma_p$.

For trace gas detection, nitrogen is the dominant buffer gas. Therefore, the overall heat-transfer characteristics of the gas mixture do not change significantly. The main variations arise from the absorption-induced heat-source power $P_Q$ and its spatial distribution $\psi_p$, which is determined by the absorption coefficient, relaxation time, and pump wavelength. Using the parameters listed in Table S2, we performed numerical simulations of the photothermal response for each of the gas species at the specified absorption line. The calculated phase modulation amplitude at different modulation frequency is presented in Fig. S14. For different gas species, the relaxation time affects the frequency response at high frequencies, while the absorption coefficient and pump wavelength determines the maximum photothermal phase modulation that is approximately flat for frequency below 10 kHz. This demonstrates that the suspended waveguide can act as an efficient platform to generate photothermal phase modulation for all the four gas species, although the magnitude of phase modulation varies due to different absorption of the gas species. These results show that the same sensing platform can be used to detect different gas species by using a pump laser tuned to the absorption line of a specific gas molecule.

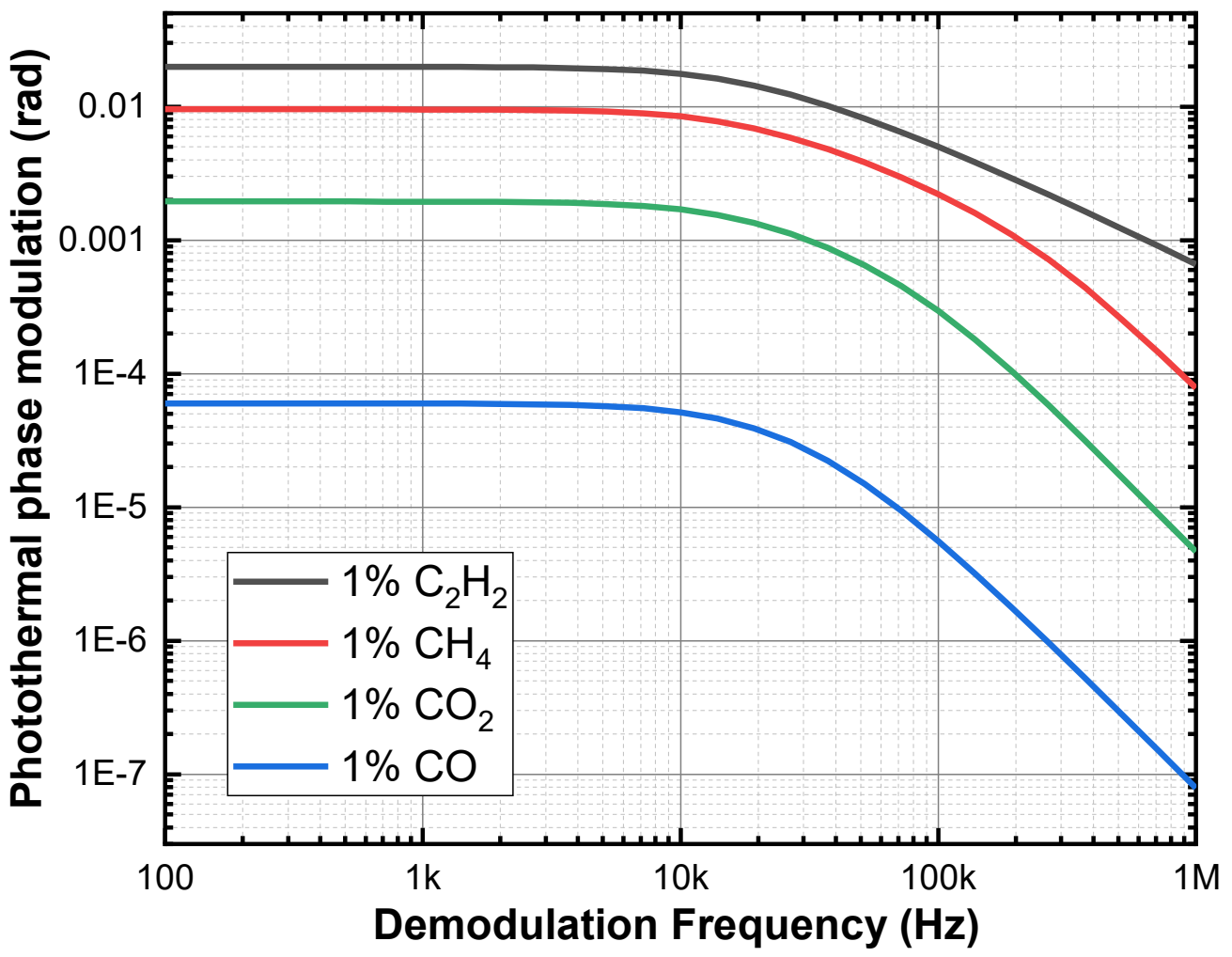

**Fig. S14** Calculated photothermal response for different gases. The gas concentrations are set to be 1% under 8 mW pump power and 1.2 cm ChGW length.

### Note 9. Performance comparison with other waveguide-based gas sensors

Detailed sensing performance between the reported state-of-the-art waveguide gas sensors are shown in Table S3. Here we use the minimum detection limit (MDL) and noise equivalent absorption (NEA) to compare the detection sensitivities. Other parameters, including waveguide types with/without suspended structure, gas confinement factor (GCF), transmission loss, waveguide length, dynamic range and response time are also listed for comparison. The demonstrated SWE-PTS sensor yields an NEA of $3.8\times10^{-7}$ cm$^{-1}$, more than 1−4 orders of magnitude higher than that of the reported waveguide sensors. The dynamic range reaches nearly 6 orders of magnitude, more than 2 orders of magnitude larger than any of the previous on-chip gas sensing techniques. These comprehensive performances pave the way for high-sensitivity, high-selectivity, large dynamic range, fast-response, and ultracompact on-chip gas sensors.

**Table S3** Performance comparisons of the reported waveguide-based gas sensors

| Method & Ref. | Gas &λ | Waveguide types | *Γ* (%) | Loss (dB/cm) | L (cm) | MDL @ time | NEA ($cm^{-1}$) | Dynamic range | Response time |
|---|---|---|---|---|---|---|---|---|---|
| DAS[3] | $CH_4$ 1650 nm | Si Rectangular | 28.3 | 2 | 10 | 100 ppm @60 s | $4.4\times10^{-5}$ | $>1.5\times10^{2}$ | >400 s |
| DAS[4] | $CH_4$ 3310 nm | ChG Rectangular | 12.5 | 8 | 0.5 | 330 ppm | $1.3\times10^{-2}$ | $>1.5\times10^{2}$ | NA |
| DAS[5] | $CH_4$ 3310 nm | ChG Rectangular | 10 | 7 | 2 | 4000 ppm | $1.6\times10^{-1}$ | $>2.5\times10^{2}$ | NA |
| DAS[6] | $CH_4$ 3270 nm | Si Slot | 69 | 8.3 | 1.15 | 0.3 ppm @50 s | $1.5\times10^{-5}$ | $>3.3\times10^{3}$ | >100 s |
| DAS[7] | $C_2H_2$ 2566 nm | $Ta_2O_5$ Suspended | 107 | 6.8 | 2 | 7 ppm @25 s | $1.8\times10^{-4}$ | $>1.4\times10^{4}$ | NA |
| DAS[8] | $CO_2$ 4240 nm | Si Suspended | 44 | 3 | 0.32 | 1000 ppm | $3.1\times10^{-1}$ | $>2.5\times10^{2}$ | 2 s |
| DAS[9] | $CH_4$ 3291 nm | ChG Suspended | 112 | 4.5 | 1 | 30.3ppm @43.4 s | $3.2\times10^{-4}$ | $>2.6\times10^{3}$ | NA |
| WMS[10] | $C_2H_2$ 1532 nm | SU8 Rectangular | 1.7 | 3 | 13 | 28.7ppm @48.6 s | $3\times10^{-5}$ | $>1.3\times10^{4}$ | 3 s |
| WMS[9] | $CH_4$ 3291 nm | ChG Suspended | 112 | 4.5 | 1 | 5.9 ppm @64.2 s | $6.2\times10^{-5}$ | $>1.3\times10^{4}$ | NA |
| WMS[11] | $CH_4$ 3291 nm | Si Rectangular | 23 | 0.71 | 2 | 78 ppm @0.2 s | $8.2\times10^{-4}$ | $>5\times10^{3}$ | 3 s |
| WMS[12] | $CH_4$ 3291 nm | ChG Rectangular | 7.8 | 1.52 | 2 | 140 ppm @32.4 s | $1.5\times10^{-3}$ | $>2.1\times10^{3}$ | NA |
| RIS[13] | $NH_3$ 1550 nm | Si MRR | NA | NA | 5 µm (radii) | 5 ppm | NA | $>1.6\times10^{2}$ | >30 s |
| PTS[14] | Resist 3419 nm | Si-MRR Suspended | NA | NA | 25µm (radii) | 5 ppm | $1.5\times10^{-4}$ | $>10^{4}$ | NA |
| PTS[15] | $C_2H_2$ 1531 nm | ChG Rectangular | 10.5 | 2.5 | 2 | 4 ppm @170 s | $4.5\times10^{-6}$ | $1.4\times10^{5}$ | NA |
| PTS[16] | $CO_2$ 2004 nm | LN Rectangular | 7.4 | 1.2 | 9.12 | 870 ppm @190 s | $1.2\times10^{-4}$ | $>6.3\times10^{2}$ | NA |
| **PTS [This]** | **$C_2H_2$ 1531 nm** | **ChG Suspended** | **95** | **2.6** | **1.2** | **330 ppb @65 s** | **$3.8\times10^{-7}$** | **$9.1\times10^{5}$** | **<1 s** |

DAS: direct absorption spectroscopy, WMS: wavelength modulation spectroscopy, a variation of DAS, RIS, reflective index spectroscopy, PTS: photothermal spectroscopy, Si: silicon, SU8: photoresist based on EPONSU-8 epoxy resin, ChG: chalcogenide glass, LN: lithium niobate, Resist: polymer photoresist AZ5214, MRR: micro-ring resonator, *Γ*: gas confinement factor, MDL: minimum detection limit with different averaging time, NEA: noise-equivalent absorption, NA: Not available.